\documentclass[twocolumn, superscriptaddress, preprintnumbers, amsmath, amssymb]{revtex4}

\allowdisplaybreaks
\allowdisplaybreaks[4]
\usepackage{CJK}
\usepackage{graphicx}
\usepackage{dcolumn}
\usepackage{bm}
\usepackage{mathptmx}
\usepackage[
            pdfstartview=FitH,
            CJKbookmarks=true,
            bookmarksnumbered=true,
            bookmarksopen=true,
            colorlinks=true,
            linkcolor=blue,
            anchorcolor=blue,
            citecolor=blue
            ]{hyperref}
\begin{document}

\title{Spin squeezed states and wobbling motion in
collective Hamiltonian}

\author{Q. B. Chen}\email[Corresponding author:~]{qbchen@phy.ecnu.edu.cn}
\affiliation{Department of Physics, East China Normal University,
Shanghai 200241, China}

\author{S. Frauendorf}\email[Corresponding author:~]{sfrauend@nd.edu}
\affiliation{Physics Department, University of Notre Dame, Notre
Dame, IN 46556, USA}

\date{\today}

\begin{abstract}

A semiclassical approach is proposed to calculate the collective potential
and mass parameters to formulate a collective Hamiltonian capable of describing
the wobbling motion in both even-even and odd-mass systems. By diagonalizing the
resulting collective Hamiltonian (CH), one can obtain the energies and wave functions
associated with the wobbling states. Furthermore, a novel technique called spin
squeezed state (SSS) maps is introduced based on the derived wave functions.
To validate the results obtained from the collective Hamiltonian, a comparative
analysis is conducted against predictions from the triaxial rotor model (TRM) and
particle triaxial rotor (PTR) model. Notably, the SSS plots determined
using the TRM and PTR models exhibit a strong correlation with the probability density
distributions of the wave functions obtained from the CH. This correlation
highlights the consistency and coherence between the different theoretical
approaches when describing the wobbling phenomenon and associated rotational
dynamics.

\end{abstract}

\maketitle


\section{Introduction}

Nuclear wobbling motion, originally proposed by Bohr and Mottelson~\cite{Bohr1975},
is a unique phenomenon observed in triaxially deformed rotating nuclei. This
motion involves the nucleus rotating around the principal axis with the
largest moment of inertia, which then executes harmonic oscillations
about the space-fixed angular momentum vector. The energy spectra associated
with this motion produce a series of rotational $E2$ bands that correspond
to different oscillation quanta ($n$). These bands are characterized by
transitions with $\Delta I=1$, which predominantly exhibit $E2$ character.

However, the existence of wobbling motion in even-even nuclei, where
no intrinsic angular momentum is involved, remains unconfirmed
experimentally. Wobbling bands have been observed in odd-mass nuclei,
where the coupling between a triaxial rotor and a high-$j$ particle or
hole can give rise to two distinct wobbling modes~\cite{Frauendorf2014PRC}.
These modes are referred to as Longitudinal Wobbling (LW) and Transverse
Wobbling (TW), depending on whether the angular momentum of the
high-$j$ particle or hole is parallel or perpendicular to the principal
axis with the largest moment of inertia. Consequently, the wobbling energy
increases with spin for LW, while it decreases for TW~\cite{Frauendorf2014PRC}.
In a subsequent work by Chen and Frauendorf~\cite{Q.B.Chen2022EPJA},
a more comprehensive classification of wobbling motion was introduced
based on the topology of classical orbits visualized using corresponding
Spin Coherent State (SCS) maps. These maps (also called as azimuthal
plots in Refs.~\cite{F.Q.Chen2017PRC, Q.B.Chen2018PRC_v1, Streck2018PRC}
when applied for the visualization of total angular momentum) provide probability
distributions for the orientation of the angular momentum on the unit
sphere projected onto the polar angle $\theta$ and azimuthal angle $\phi$
plane. According to this classification, LW corresponds to a revolution of
the total angular momentum $\bm{J}$ around the axis with the largest
moment of inertia, whereas TW corresponds to a revolution of $\bm{J}$
around an axis perpendicular to the axis with the largest moment of
inertia. By employing these scientific classifications and models,
one has gained a deeper understanding of the wobbling motions
exhibited by triaxial rotors coupled with high-$j$ quasiparticles.

Currently, the experimental observations of wobbling bands are primarily
reported in odd-proton nuclei. Notable examples include
$^{161}$Lu~\cite{Bringel2005EPJA}, $^{163}$Lu~\cite{Odegaard2001PRL, Jensen2002PRL}, $^{165}$Lu~\cite{Schonwasser2003PLB}, $^{167}$Lu~\cite{Amro2003PLB},
$^{167}$Ta~\cite{Hartley2009PRC}, and the most recent discovery in
$^{151}$Eu~\cite{Mukherjee2023PRC} in the $A\approx 160$ mass region,
$^{135}$Pr~\cite{Matta2015PRL, Sensharma2019PLB} and $^{133}$La~\cite{Biswas2019EPJA}
in the $A\approx 130$ mass region, and $^{187}$Au~\cite{Sensharma2020PRL} and
$^{183}$Au~\cite{Nandi2020PRL} in the $A\approx 190$ mass region.
Additionally, wobbling bands have been observed in odd-neutron nuclei
such as $^{105}$Pd~\cite{Timar2019PRL}, $^{127}$Xe~\cite{Chakraborty2020PLB},
and $^{133}$Ba~\cite{Devi2021PLB}. More recently, wobbling bands
have also been pointed out in even-even nuclei with a two-quasiparticle
configuration, specifically in $^{130}$Ba~\cite{Petrache2019PLB, Q.B.Chen2019PRC_v1,
Y.K.Wang2020PLB} and $^{136}$Nd~\cite{F.Q.Chen2021PRC, B.F.Lv2022PRC}. It is important
to acknowledge that certain proposed wobblers continue to be a topic of
controversy~\cite{Frauendorf2018PRC, Tanabe2018PRC, Lawrie2020PRC, B.F.Lv2021PRC,
Tanabe2017PRC, B.F.Lv2022PLB, S.Guo2022PLB, Nomura2022PRC}. Nevertheless, these
experimental observations contribute to the scientific understanding
of wobbling motion in various nuclear systems.

On the theoretical front, the wobbling motion was initially proposed within the
framework of the Triaxial Rotor Model (TRM)~\cite{Bohr1975}. The TRM allows
for an investigation of the angular momentum geometry associated with the
wobbling motion in a triaxially deformed even-even nucleus~\cite{W.X.Shi2015CPC,
B.Qi2021JPG, Q.B.Chen2022EPJA}. Subsequent to the discovery of the first wobbling
structure in the odd-proton nucleus $^{163}\rm Lu$~\cite{Odegaard2001PRL}, the
Particle-Triaxial-Rotor (PTR) model~\cite{Hamamoto2002PRC, Hamamoto2003PRC,
Frauendorf2014PRC, Streck2018PRC, Q.B.Chen2019PRC_v1, Q.B.Chen2020PLB_v1,
Broocks2021EPJA, L.Hu2021PRC, Q.B.Chen2022EPJA, H.Zhang2022PRC, H.M.Dai2023PRC}
and its approximate solutions~\cite{Raduta2017PRC, Budaca2018PRC,
Budaca2021PRC, Budaca2022PRC_v1} have been employed to describe the wobbling mode.

In addition, various efforts have been made to extend the cranking model within
the mean field theory framework in order to study the wobbling motion. However,
due to the mean field approximation, the cranking model only yields the yrast
sequence for a given configuration. To describe wobbling excitations,
it is necessary to go beyond the mean field approximation. Currently,
this has been achieved by incorporating quantum correlations through methods
such as the random phase approximation (RPA)~\cite{Shimizu1995NPA,
Matsuzaki2002PRC, Matsuzaki2004PRCa, Matsuzaki2004PRC, Shimizu2008PRC,
Shoji2009PTP, Frauendorf2015PRC}, the angular momentum
projection (AMP) method~\cite{Shimada2018PRC, Y.K.Wang2020PLB,
F.Q.Chen2021PRC}, and the collective Hamiltonian (CH) method~\cite{Q.B.Chen2014PRC,
Q.B.Chen2016PRC_v1}. These theories offer means to incorporate quantum
effects beyond the mean field level in the study of wobbling motion.

The orientation of a nucleus in the rotating mean field is typically described
using spherical coordinates, namely the polar angle $\theta$ and azimuth
angle $\phi$. In the collective Hamiltonian framework for wobbling
modes~\cite{Q.B.Chen2014PRC, Q.B.Chen2016PRC_v1}, the azimuth angle
$\phi$ is adopted as the collective coordinate due to the relatively
easier motion along the $\phi$ direction than in the $\theta$ direction.
The quantum fluctuations along $\phi$ are carefully taken into account
to go beyond the standard mean field approximation. This model has been
employed to systematically study the triaxial rotor, transverse, and
longitudinal wobblers, and confirm the variation trends of their wobbling
energies~\cite{Q.B.Chen2014PRC}. Furthermore, the collective
Hamiltonian has been applied to investigate the wobbling motion
in $^{135}$Pr~\cite{Q.B.Chen2016PRC_v1}, with results clearly
indicating that the experimental energy spectra of both yrast
and wobbling bands are well reproduced by the collective
Hamiltonian. The mechanism involved in the transition from TW to LW
is elucidated through investigations of the softness and shapes of
the collective potential. Notably, the collective Hamiltonian
framework has also achieved considerable success in describing
chiral motion~\cite{Q.B.Chen2013PRC, Q.B.Chen2016PRC, X.H.Wu2018PRC},
which is also the fingerprint of the triaxial deformation.

It is worth noting that the employment of the Titled Axis Cranking (TAC)
approximation in the collective Hamiltonian approach, as described in
Refs.~\cite{Q.B.Chen2014PRC, Q.B.Chen2016PRC_v1, Q.B.Chen2013PRC, Q.B.Chen2016PRC,
X.H.Wu2018PRC}, leads to a situation where the angular momentum is not a
good quantum number. Instead, it is evaluated as the expectation value of
the angular momentum operator on the cranking state. To address this limitation,
the present paper proposes a novel approach, in which both the collective potential
and mass parameter of collective Hamiltonian are derived by means of a
semiclassical approach, which is later re-quantized. This approach provides
a framework where the total angular momentum retains its nature as a good
quantum number. The derived collective Hamiltonian is subsequently employed
to investigate the wobbling motion in both even-even and odd-mass PTR systems
in order to demonstrate its validity and limitations. The collective
Hamiltonian provides an instructive perspective on the results of
the PTR calculations in terms of the familiar ``potential + kinetic"
energy paradigm. In addition, it represents  an extension of the microscopic
TAC mean field approach for including the collective rotation mode about
all three principal axes of the triaxial nucleus. Our future work will
explore this avenue.

Furthermore, to enable a comprehensive comparison between the wave functions
derived using different models, this work will introduce the concept of Spin
Squeezed States (SSS). Squeezed states, initially introduced in the domain of
quantum optics~\cite{Kitagawa1993PRA}, represent a generalization of coherent
states~\cite{Klauder1985book}. In contrast to coherent states, which minimize
the uncertainty product $\Delta x \Delta p$, squeezed states exhibit a smaller uncertainty
in either $\Delta x$ or $\Delta p$ at the expense of a larger
complementary width. The SSS states provide a valuable interpretation
from a quantum mechanical perspective as they connect the discrete
$\bm{J}$-space representation with the continuous coordinate $\phi$'s wave
function. This approach allows for a deeper understanding of the
wobbling motion within the context of the collective Hamiltonian
framework, providing insights that go beyond the scope of
traditional analysis.


\section{Theoretical framework}

\subsection{Spin squeezed states}

The Hilbert space of the quantal system with one degree of freedom and
good absolute angular momentum $j$ is spanned by the so-called ``$k$-states"
$|k\rangle$, which possess a projection $-j\leq k\leq j$ on the quantization
3 axis. The motion of the system is subject to the constraint imposed by the
conservation of angular momentum, which restricts it to the sphere of
constant angular momentum
\begin{align}\label{eq:jsphere}
 \bm{j}^2=j_1^2+j_2^2+j_3^2=j(j+1).
\end{align}

In the context of this quantal system, it is possible to identify the
angular momentum projection operator $\hat{j}_3$ as the momentum operator
$\hat{p}$, which represents the momentum along a particular direction.
Consequently, the angle operator $\hat{\phi}$, responsible for determining
the orientation of the angular momentum vector $\hat{\bm{j}}$ projection
in the 1-2 plane, can be interpreted as the conjugate position operator
$\hat{q}$. This interpretation allows us to establish a correspondence
between the quantum mechanical operators and their classical counterparts.
In particular, we find that the following commutation relation
\begin{align}\label{eq:can}
 [\hat p,\hat q]=[\hat j_3,\hat\phi]=-i.
\end{align}
In the classical counterpart of the system, the quantities $j_3$ and $\phi$
correspond to the canonical variables $p$ and $q$, respectively.

In the following, we introduce the Spin Squeezed States (SSS). The over-complete,
non-orthogonal set of SSS are expressed in terms of $k$-states as
\begin{align}\label{eq:squeezed1}
 \vert \phi \rangle=\frac{1}{\sqrt{2j+1}}\sum\limits_{k=-j}^je^{i\phi k}\vert k\rangle
 ={\cal R}_3 ^\dagger(\phi)\vert \phi=0\rangle,
\end{align}
which are generated by rotating the state $|\phi=0\rangle$
state about the 3 axis within the range  $-\pi\leq\phi\leq\pi$.
Note that in the case of half-integer $j$, the spinor $\vert k\rangle$
obeys ${\cal R}_3^\dagger(2\pi)\vert k\rangle=-\vert k\rangle$, which
implies $\vert \phi +2\pi\rangle=-\vert \phi \rangle$. For this reason
the explicit definition of the $\phi$ range is of importance.
The SSS states are normalized but non-orthogonal,
\begin{align}\label{eq:overlapSSS}
  \vert \langle \phi\vert \phi'\rangle\vert=\left\{
  \begin{array}{cl}
1, & \textrm{if}~~\phi=\phi',\\
\\
\left|\displaystyle \frac{\sin\left[(j+1/2)(\phi-\phi')\right]}
 {(2j+1)\sin[(\phi-\phi^\prime)/2]}\right|, & \textrm{if}~~\phi\neq\phi'.
  \end{array}
\right.
\end{align}
The overlap $\vert\langle \phi_n\vert \phi'\rangle\vert \approx
0.64$ at $\phi'=(\phi_n+\phi_{n+1})/2$ and $=0$ at
$\phi'=\phi_{n+1}$.

The SSS set is massive over complete, because the dimension of the
$\bm{j}$-space is $2j+1$ whereas the dimension of the SSS space is
infinite. There are infinite many possible transformations from the
SSS basis back to the $k$-basis. One obvious transformation is
\begin{align}\label{eq:SSStoIK}
 \vert j k\rangle=\frac{\sqrt{2j+1}}{2\pi}\int_{-\pi}^{\pi}
  d\phi ~e^{-i \phi k}\vert \phi\rangle,
\end{align}
which corresponds to the resolution of the identity operator
\begin{align}\label{eq:SCSto1_1}
 \hat 1
 =\sum_k\vert jk\rangle\langle jk\vert
 =\frac{2j+1}{2\pi}\int_{-\pi}^{\pi}
 d\phi~\vert  \phi\rangle\langle  \phi \vert.
\end{align}

The probability distribution of the SSS is given by
\begin{align}\label{eq:PSSS}
 P(\phi)_{\nu}= \frac{1}{2 \pi}
  \sum\limits_{K,K'=-I}^Ie^{-i(K-K')\phi}\rho^{(\nu)}_{KK'},
\end{align}
where $\phi$ is the angle with the short axis in the short-medium axis plane.
The $\rho^{(\nu)}_{KK'}$ is the density matrix or the reduced density matrix
composed by the expansion coefficient of the eigen-function on the basis~\cite{Q.B.Chen2022EPJA}.
In the framework of TRM and the collective Hamiltonian introduced subsequently
in Sec.~\ref{sub:col}, the coefficients are denoted as $C_{IK}^{(\nu)}$,
and $\rho^{(\nu)}_{KK'}$ is calculated by~\cite{Q.B.Chen2022EPJA}
\begin{align}\label{eq9rho}
  \rho^{(\nu)}_{KK'}=C_{IK}^{(\nu)}C_{IK'}^{(\nu)*}.
\end{align}
In the PTR model with the coefficients $C_{IKk}^{(\nu)}$ (with $k$
here being the projection of the particle angular momentum on the
quantizated 3 axis), $\rho^{(\nu)}_{KK'}$ is the reduced density
matrix~\cite{Q.B.Chen2022EPJA}
\begin{align}
  \rho^{(\nu)}_{KK'}=\sum_{k}C_{IKk}^{(\nu)}C_{IK'k}^{(\nu)*},
\end{align}
which is constructed by averaging over the degrees of freedom $k$ that
are not of interest for the moment.

The SSS probability fulfills the normalization condition as
\begin{align}
  \int_{-\pi}^\pi P(\phi)_{\nu}~d\phi=1.
\end{align}

A complete orthogonal basis can be also spanned by the discrete
``$\phi$-states" $\vert \phi_n\rangle$, which are the SSS taken at
the discrete angles $\phi_n =\frac{2\pi (n-j)}{2j+1}$, $n=0$, ...,
$2j$ (c.f. Eq.~(\ref{eq:overlapSSS})). The deviation of these ``angle
states" from being eigenstates of $e^{i\hat \phi}$ is of the order
$1/2j$. They are localized in $\phi$ around $\phi_n$. The SSS
interpolate between the discrete $\phi$-states, which provides the
information about the phase relations between the discrete
$\phi$-states.

\begin{figure}[ht]
 \includegraphics[width=7.5 cm]{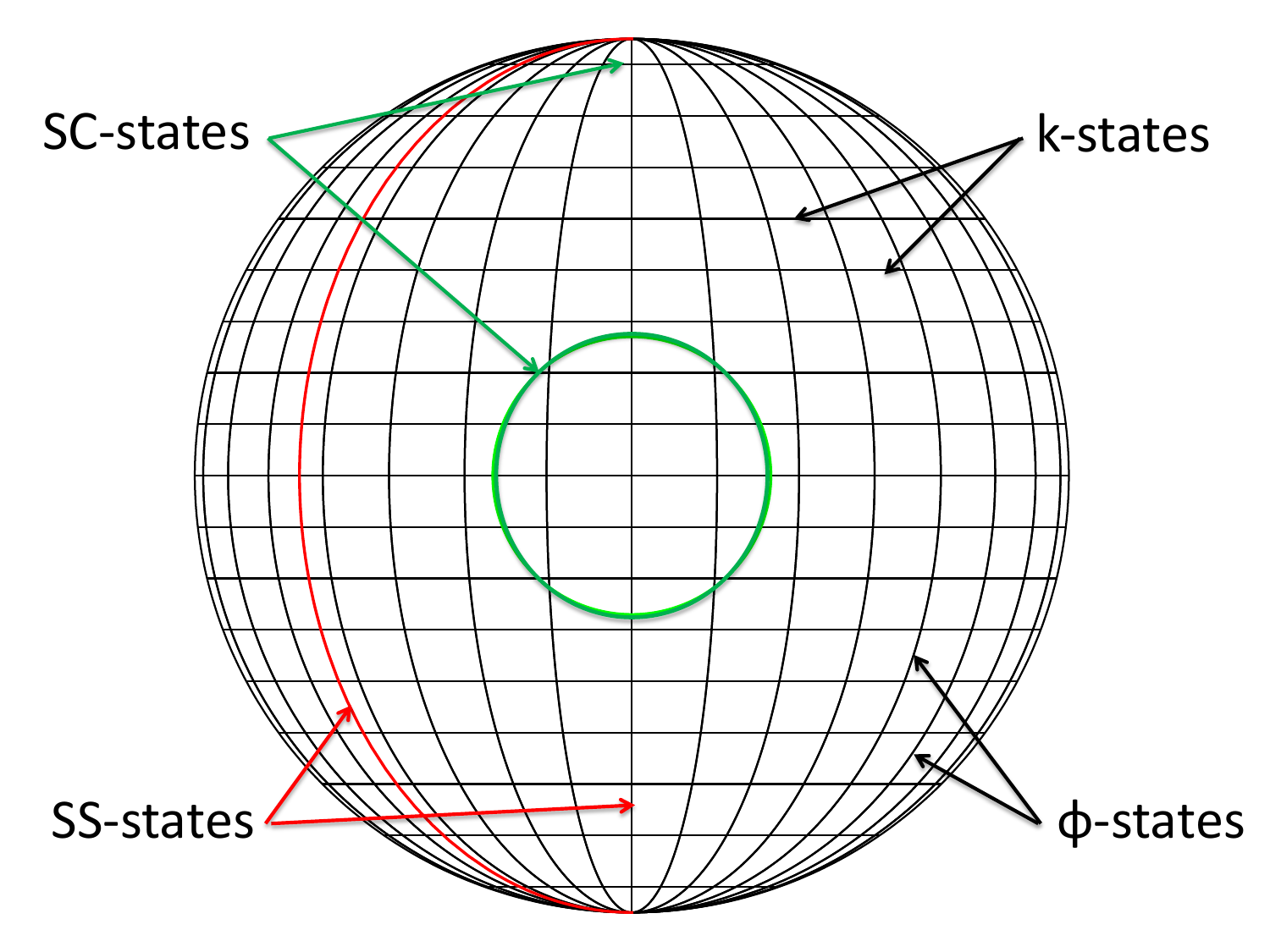}
 \caption{\label{f:StateClass}  Schematic illustration of the different
 states used as basis sets in the $\bm{j}$-space. Each state is represented
 by a curve along which the endpoint of angular momentum vector is distributed
 with equal probability. The $k$-states are circles of constant latitude $k$.
 The $\phi$-states are semicircles of constant $\phi$. The figure displays
 the pertaining meridians. For the continuous set of SCS only the
 states $\vert \theta=0,\phi=0\rangle$ and $\vert \theta=\pi/2,
 \phi=0\rangle$ are examples shown. For the continuous set of SSS
 only the states $\vert \phi=0\rangle$ and $\vert \phi=-\pi/4\rangle$
 are shown.}
\end{figure}

Figure ~\ref{f:StateClass} illustrates the basis sets of $k$-state,
$\phi$-state, SSS, as well as SCS introduced in Ref.~\cite{Q.B.Chen2022EPJA}.
The set of $k$-states is orthonormal and complete. The $j_3$ takes
the sharp values $k$ while $\phi$ is uniformly distributed
over the circles of constant latitude, which is often illustrated in form of a
precession cones. For the $\phi$-states, $\hat \phi$  is localized
at the discrete values $\phi_n$ while $j_3$ is uniformly distributed
over the semicircles of constant longitude, which can be imagined as
a semicircular sheets. They represent the momentum and coordinate
representations of the states in the $\bm{j}$-space. As mentioned above,
the SSS states interpolate between the discrete $\phi$-states,
facilitating the extraction of phase relations among these discrete
states. The set of the SCS is generated by rotating the generating
state $|j, k=j\rangle$ over the whole angular momentum
sphere~\cite{Q.B.Chen2022EPJA}.

The notation squeezed states comes from quantum optics. They are a
generalization of coherent states. For coherent states the
uncertainty product $\Delta x \Delta p$ is minimized. For squeezed
states either $\Delta x$ or $\Delta p$ is smaller than for the
optimal coherent states (and the complementary width is larger). The
SSS uncertainty $\Delta \phi\approx 3/(2j)$ (see
Eq.~(\ref{eq:overlapSSS})) is smaller than the SCS width of
$1/\sqrt{2j}$ because $k$ has a large width of $2j+1$.

The proposed SSS plots will be applied to the triaxial rotor and
particle-rotor systems in this work in combination with the collective
Hamiltonian introduced below.

\subsection{Triaxial rotor model in angle-momentum representation}
\label{sub:col}

The classical mechanics of gyroscopes offers a framework for classifying
quantal states. In this context, the classical orbits of the angular
momentum vector, relative to the body-fixed principal axes, represent
the intersection lines between the sphere of constant angular momentum
and the ellipsoid of constant energy. Detailed discussions on the
construction of these orbits can be found in Refs.~\cite{Frauendorf2018PS,
Frauendorf2014PRC, Q.B.Chen2022EPJA}. For the triaxial rotor the orbits
are given by the two angles $\theta$ and
$\phi$ as
\begin{align}
\label{eq4}
& J_1={\sqrt{I(I+1)}}\sin \theta\cos \phi,\\
& J_2={\sqrt{I(I+1)}}\sin \theta\sin \phi,\\
\label{eq5}
& J_3={\sqrt{I(I+1)}}\cos\theta,
\end{align}
which are restricted to the sphere of constant angular momentum
\begin{align}
 \bm{J}^2=J_1^2+J_2^2+J_3^2=I(I+1).
\end{align}
Namely, $\theta$ is the angle between the angular momentum $\bm{J}$ and
the 3 axis and $\phi$ is the angle between the projection of $\bm{J}$
onto the 1-2 plane and the 1 axis.

With the above definitions, the Triaxial Rotor Model (TRM) Hamiltonian~\cite{Bohr1975}
\begin{align}\label{HTRM-J}
 H_{\textrm{TR}}
  &=\frac{\hat J_1^2}{2\mathcal{J}_1}+\frac{\hat J_2^2}{2\mathcal{J}_2}
   +\frac{\hat J_3^2}{2\mathcal{J}_3}
 \end{align}
 can be rewritten as a classical Hamiltonian
 \begin{align}
  H_{\textrm{TR}}=I(I+1)\Big(\frac{\sin^2\theta\cos^2\phi}{2\mathcal{J}_1}
   +\frac{\sin^2\theta\sin^2\phi}{2\mathcal{J}_2}
  +\frac{\cos^2\theta}{2\mathcal{J}_3}\Big),
  \end{align}
  which can be cast into the form
\begin{align}\label{HTRM-f}
  H_{\textrm{TR}}=V(\phi)+\frac{J_3^2}{2B(\phi)},
\end{align}
with
\begin{align}
\label{VTRM}
 V(\phi)&=I(I+1)\Big(\frac{\cos^2\phi}{2\mathcal{J}_1}
  +\frac{\sin^2\phi}{2\mathcal{J}_2}\Big),\\
\label{BTRM}
 B(\phi)&=\frac{1}{2}\left[\frac{1}{2\mathcal{J}_3}
  -\frac{\cos^2\phi}{2\mathcal{J}_1}
  -\frac{\sin^2\phi}{2\mathcal{J}_2}\right]^{-1}.
\end{align}
In the above formula, the $\mathcal{J}_{1,2,3}$ are the corresponding
moments of inertia of the three principal axes. It is crucial to
highlight that the derivations presented above for the
collective potential $V(\phi)$ and mass parameter $B(\phi)$ are carried
out without any approximation. This rigorous treatment allows for a
comprehensive and precise characterization of the quantum behavior
associated with wobbling motion in triaxial nuclei.

The classical TRM Hamiltonian (\ref{HTRM-f}) in $J_3$-$\phi$
representation has the form of a kinetic term with the $\phi$-dependent inertia
parameter $B(\phi)$ and the periodic potential $V(\phi)$ as given by Eqs.~(\ref{BTRM})
and (\ref{VTRM}). The step of quantization of the classical energy is not unique.
To quantize it, we apply the general Pauli prescription~\cite{Podolsky1928PR, Pauli1933}.
The quantized form of the collective Hamiltonian (CH) reads as~\cite{Q.B.Chen2013PRC,
Q.B.Chen2014PRC, Q.B.Chen2016PRC_v1}
\begin{align}\label{eq2}
 H_{\textrm{CH}}=-\frac{\hbar^2}{2\sqrt{B(\phi)}}\frac{\partial}{\partial \phi}
  \frac{1}{\sqrt{B(\phi)}}\frac{\partial}{\partial \phi}+V(\phi).
\end{align}
The volume element in the present collective space is
\begin{align}
 \int d\tau_{\textrm{CH}}=\int_{-\pi}^\pi d\phi \sqrt{B(\phi)}.
\end{align}

The CH Hamiltonian is diagonalized in the basis
\begin{align}\label{TRM-basis}
 |\psi_K\rangle=\frac{1}{\sqrt{2\pi}}\frac{\exp[iK\phi]}{B^{1/4}(\phi)}.
\end{align}
The calculations of the CH matrix elements
 is straightforward. It is worth pointing out that the
derivation of $\partial/\partial \phi$ should act to the mass
parameter term $B^{1/4}(\phi)$ in the denominator of Eq.~(\ref{TRM-basis}).
The results of the diagonalization in the basis $-I \leq K\leq I$ agrees
exactly with the diagonalization of the original TRM Hamiltonian (\ref{HTRM-J}).
It provides all solution with the four representations of the D$_2$ group
$\mathcal{R}_1(\pi)=\pm 1$ and $\mathcal{R}_3(\pi)=\pm 1$. The TRM describes the
rotational excitations built on the ground state even-even nuclei. Therefore the
intrinsic state is the quasiparticle vacuum, which has the symmetry
$\mathcal{R}_1(\pi)=\mathcal{R}_3(\pi)= 1$. The TRM rotational states must have
the same symmetry, because a rearrangement of the principal axes must leave the
product of the intrinsic and rotational wave function invariant (see Ref.~\cite{Bohr1975}).
The Hamiltonian couples only basis states $K$ with $K\pm 2$ and is invariant with respect
to $K\rightarrow -K$. The completely symmetric representation corresponds to even-$K$ and
symmetric solutions for even-$I$ and anti-symmetric solutions for odd-$I$.

The collective function is written as
\begin{align}\label{eq8}
  |\Psi\rangle_\nu=\sum_K C_{IK}^{(\nu)}|\psi_K\rangle,
\end{align}
in which the expansion coefficients $C_{IK}^{(\nu)}$ are obtained by
diagonalizing the CH Hamiltonian (\ref{eq2}) and used to construct
the density matrix (\ref{eq9rho}) to calculate the SSS plots. Additionally we
diagonalized it in an extended $K$-basis, large enough that the results
did not depend on the cut-off.

\subsection{Collective Hamiltonian for the Particle-Triaxial-Rotor Model}
\label{sub2}

The Particle-Triaxial-Rotor (PTR) model couples  high-$j$ particles to the triaxial rotor
core. The corresponding Hamiltonian has been given in the textbook, e.g.,
Ref.~\cite{Bohr1975}
\begin{align}\label{eq:HPTR}
 \hat{H}_{\textrm{PTR}}=\hat{h}_p
  +\sum_{i=1,2,3} A_i(\hat{J}_i^2-2\hat{J}_i\hat{j}_i+\hat{j}_i^2).
\end{align}
with the inertial parameters $A_i=1/(2\mathcal{J}_i)$. The
particle Hamiltonian $\hat{h}_p$ is taken as single-$j$ shell Hamiltonian
\begin{align}\label{eq:hproton}
 \hat{h}_p&=\kappa\left[\left(3\hat{j}_3^2-\bm{j}^2\right)\cos\gamma
     +\sqrt3\left(\hat{j}_1^2-\hat{j}_2^2\right)\sin\gamma\right].
\end{align}
The angle $\gamma$ serves as the triaxial deformation parameter and the
coupling parameter $\kappa$ is proportional to the quadrupole deformation
parameter $\beta$.

As discussed in our previous paper \cite{Q.B.Chen2022EPJA},
the PTR Hamiltonian is diagonalized in the product basis
$\vert IIK\rangle\vert jk\rangle$, where $\vert IIK\rangle$ are the
rotational states for half-integer $I$ and $\vert jk\rangle$ the high-$j$
particle states in good spin $j$ approximation. The eigenstates are
hence expressed as
\begin{equation}
 |II\nu\rangle=\sum_{K,k} C_{IKk}^{(\nu)} \vert IIK\rangle\vert jk\rangle.
\end{equation}
The coefficients $C_{IKk}^{(\nu)}$ of the states in the triaxially
deformed odd-$A$ nuclei are not completely free. They are restricted by
requirement that collective rotor states must be symmetric
representations of the D$_2$ point group. When the $K$ and $k$
in the sum run respectively from $-I$ to $I$ and from $-j$ to $j$,
their difference $K-k$ must be even, and one half of all coefficients
is fixed by the symmetry relation
\begin{align}\label{eq:D2symmetry}
C_{I-K-k}^{(\nu)}=(-1)^{I-j}C_{IKk}^{(\nu)}.
\end{align}
From the amplitudes of the eigenstates $C_{IKk}^{(\nu)}$, the
reduced density matrices
\begin{align}\label{eq:TWrhoj}
 \rho_{kk'}^{(\nu)}=\sum_K C_{IKk}^{(\nu)}C_{IKk'}^{(\nu)*}
\end{align}
and
\begin{align}\label{eq:TWrhoJ}
 \rho_{KK'}^{(\nu)}=\sum_k C_{IKk}^{(\nu)}C_{IK'k}^{(\nu)*}
\end{align}
are obtained, which contain the information about the
particle angular momentum $\bm{j}$ and the total angular momentum $\bm{J}$,
respectively. The respective reduced density matrices are
plugged into Eq.~(\ref{eq:PSSS}) to generate the pertaining SSS plots.
We used the same method in our preceding paper \cite{Q.B.Chen2022EPJA}
to generate the SCS plots.

In order to construct a CH we apply the adiabatic approximation.
The particle angular momentum $\hat{j}_i$ is treated as an operator
while the total angular momentum is treated as a classical vector as
in Eqs.~(\ref{eq4})-(\ref{eq5}). The corresponding adiabatic Hamiltonian
is written as
\begin{align}\label{eq7}
 \hat{H}_{\textrm{ad}}=\hat{h}_p+\sum_{i=1,2,3} A_i(J_i^2-2J_i\hat{j}_i+\hat{j}_i^2),
\end{align}
with $J_i$ being numbers given in Eqs.~(\ref{eq4})-(\ref{eq5})
and the inertial parameters $A_i=1/(2\mathcal{J}_i)$.
 The $\hat{H}_{\textrm{ad}}$ is diagonalized in the $\bm{j}$-space
$|jk\rangle$ (with $k$ being the 3 axis component of the particle
angular momentum $\bm{j}$ in the intrinsic frame). This indicates
that the particle angular momentum operators $\hat{j}_i$ and $\hat{j}_i^2$
are treated exactly. The adiabatic approximation is similar but slightly
different from the Tilted Axis Cranking (TAC) mean field approximation~\cite{Frauendorf1993NPA,
Frauendorf2000NPA, Q.B.Chen2014PRC, Q.B.Chen2016PRC}. For the adiabatic
PTR the recoil term $\langle \hat{j}_i^2 \rangle$ is diagonalized whereas
is  replaced by $\langle \hat{j}_i\rangle^2$ under the TAC mean field approximation.

The expectation value of the adiabatic Hamiltonian can be expressed in terms
of the polar angle $\theta$ and azimuthal angle $\phi$ of the total angular
momentum $\bm{J}$ as follows
\begin{widetext}
\begin{align}
  E_{\textrm{ad}}(\theta,\phi) &=\langle \hat{H}_{\textrm{ad}} \rangle \notag\\
  &=\langle \hat{h}_p \rangle +\sum_{i=1,2,3} A_i(J_i^2-2J_i\langle\hat{j}_i\rangle
  +\langle \hat{j}_i^2\rangle)\notag\\
  &=\langle \hat{h}_p \rangle +\sum_{i=1,2,3} A_i\langle \hat{j}_i^2\rangle
   +I(I+1)\Big[(A_1\cos^2\phi+A_2\sin^2\phi)
   +\cos^2\theta(A_3-A_1\cos^2\phi-A_2\sin^2\phi)\Big]\notag\\
  &\quad -2\sqrt{I(I+1)}\Big[A_1\sqrt{1-\cos^2\theta}\cos\phi\langle \hat{j}_1\rangle
   +A_2\sqrt{1-\cos^2\theta}\sin\phi\langle \hat{j}_2\rangle
   +A_3\cos\theta\langle \hat{j}_3\rangle\Big]\label{eq:Ead}.
\end{align}
\end{widetext}
We consider the Transverse Wobbling (TW) case. The adiabatic
energy has a minimum at $\theta=\pi/2$, because the long axis (3 axis)
of the triaxial deformed nucleus has the smallest moment of inertia and the particles
tend to align their angular momentum with the short axis (1-axis).
In order to construct a CH of the form (\ref{HTRM-f}), we expand it with respect
to $\cos\theta$, which is a small quantity near $\theta=\pi/2$.
The linear term  is zero because the expansion is around the minimum.
Further  employing the  approximation
\begin{align}\label{eq7a}
\sqrt{1-\cos^2\theta}\approx 1-\frac{1}{2}\cos^2\theta,
\end{align}
one can approximate the adiabatic energy as
\begin{widetext}
\begin{align}\label{eq6}
E_{\textrm{ad}}(\theta,\phi)
  &\approx \langle \hat{h}_p\rangle+\sum_{i=1,2,3} A_i\langle \hat{j}_i^2\rangle
   +I(I+1)(A_1\cos^2\phi+A_2\sin^2\phi)
   -2\sqrt{I(I+1)}\Big(A_1\cos\phi\langle \hat{j}_1\rangle
   +A_2\sin\phi\langle \hat{j}_2\rangle\Big)
   -2A_3J_3 \langle \hat{j}_3\rangle\notag\\
  &\quad+J_3^2\Big[(A_3-A_1\cos^2\phi-A_2\sin^2\phi)
        +\frac{1}{\sqrt{I(I+1)}}(A_1\cos\phi\langle \hat{j}_1\rangle
   +A_2\sin\phi\langle \hat{j}_2\rangle)\Big]\notag\\
  &= V(\phi)+\frac{J_3^2}{2B(\phi)},
\end{align}
\end{widetext}
where we have used Eqs.~(\ref{eq4})-(\ref{eq5}).

The collective potential $V(\phi)$
is energy $E_{\textrm{ad}}(\theta=\pi/2,\phi)$.
To obtain the mass parameter we calculate the energy variation generated
by the very small shift $\delta= \theta-\pi/2$
\begin{align}
 \Delta E
  &=E_{\textrm{ad}}(\pi/2+\delta,\phi)
  -E_{\textrm{ad}}(\pi/2,\phi)
\end{align}
Correspondingly, the angular momentum of the third component
varies as
\begin{align}
 \Delta J_3^2
  &=J_3^2(\pi/2+\delta,\phi)
   -J_3^2(\pi/2,\phi).
\end{align}
From the relationship
\begin{align}
 \Delta E=\frac{\Delta J_3^2}{2B(\phi)},
\end{align}
one obtains
\begin{align}\label{eq12}
 B(\phi)=\frac{\Delta J_3^2}{2\Delta E}.
\end{align}
One can easily check that using this method, the same mass parameter
(\ref{BTRM}) of the TRM can be obtained.


\section{Results and discussion}

\subsection{Triaxial rotor system}\label{s:TR}

For the triaxial rotor, we discuss the system with the moments
of inertia $\mathcal{J}_{m,s,l}$=30, 10, 5~$\hbar^2$/MeV for
the medium ($m$), short ($s$), and long ($l$) axes, which was
studied previously in Refs.~\cite{Frauendorf2014PRC, W.X.Shi2015CPC,
Q.B.Chen2022EPJA}. In Refs.~\cite{W.X.Shi2015CPC, Q.B.Chen2022EPJA},
the energy spectra as function of spin and the angular momentum
structure were illustrated in the framework of TRM. In addition,
Ref.~\cite{Q.B.Chen2022EPJA} introduced the two-dimensional SCS maps,
i.e., probability distribution for the orientation of the angular
momentum on the $(\theta,\phi)$ plane, to extract the classical
mechanics underpinning of the quantal triaxial rotor model from the
numerical results. In this subsection, we will focus on discussing
the information provided by the collective Hamiltonian and the
one-dimensional SSS.

\begin{figure}[!ht]
  \begin{center}
    \includegraphics[width=7.0 cm]{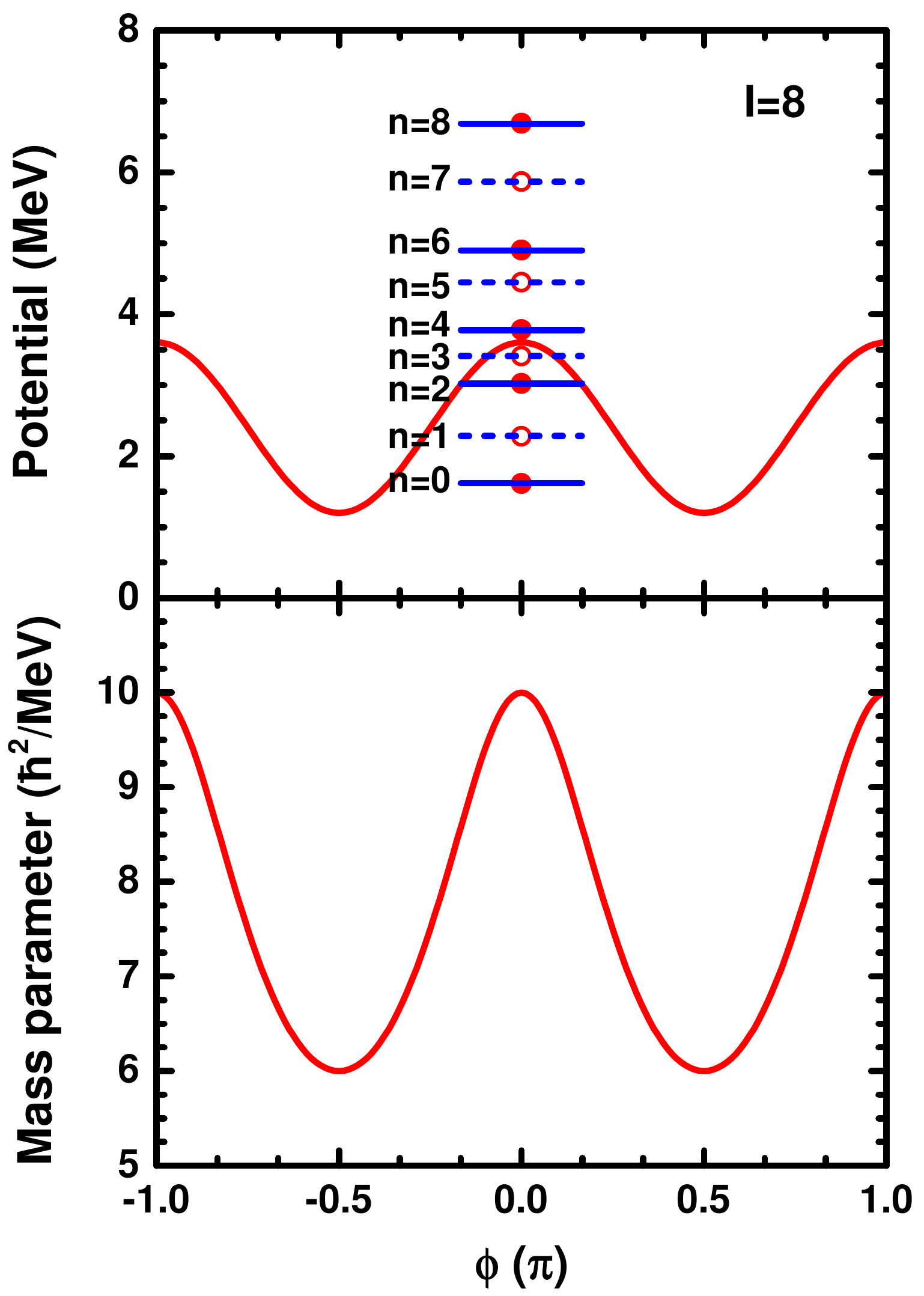}
    \caption{Upper panel: classical potential energy $V(\phi)$ as a
    function of $\phi$ for spin $I=8$. The horizontal lines show the
    quantal energies $E_\nu$ calculated by TRM, while the dots represent
    the energies obtained by the collective Hamiltonian $H_{\textrm{CH}}$,
    which incorporates a $\phi$-dependent mass parameters $B(\phi)$ shown
    in the lower panel. Note that for the even phonon states, they are
    calculated from $I=8$ states, while for the odd phonon states, they
    are calculated as the $72/90$ of $I=9$ energies. }\label{f:Energy_TR}
  \end{center}
\end{figure}

The calculated collective potential $V(\phi)$ and the collective mass
parameter $B(\phi)$ for $I=8$ are shown in Fig.~\ref{f:Energy_TR}.
Note that $\phi$ is the angle between the projection of $\bm{J}$
onto the $s$-$m$ plane and the $s$ axis. The $V(\phi)$ and $B(\phi)$ are
symmetrical with respect to $\phi=0$ line, as a result of the invariance
of the intrinsic quadrupole moments with respect to the $\textrm{D}_2$ symmetry
for a triaxial rotor system. The minima in the $V(\phi)$ and $B(\phi)$
locate at $\phi=\pm \pi/2$, which corresponds to uniform rotation about
the $m$ axis with the largest moment of inertia.

\begin{figure}[!th]
  \begin{center}
    \includegraphics[width=8.5 cm]{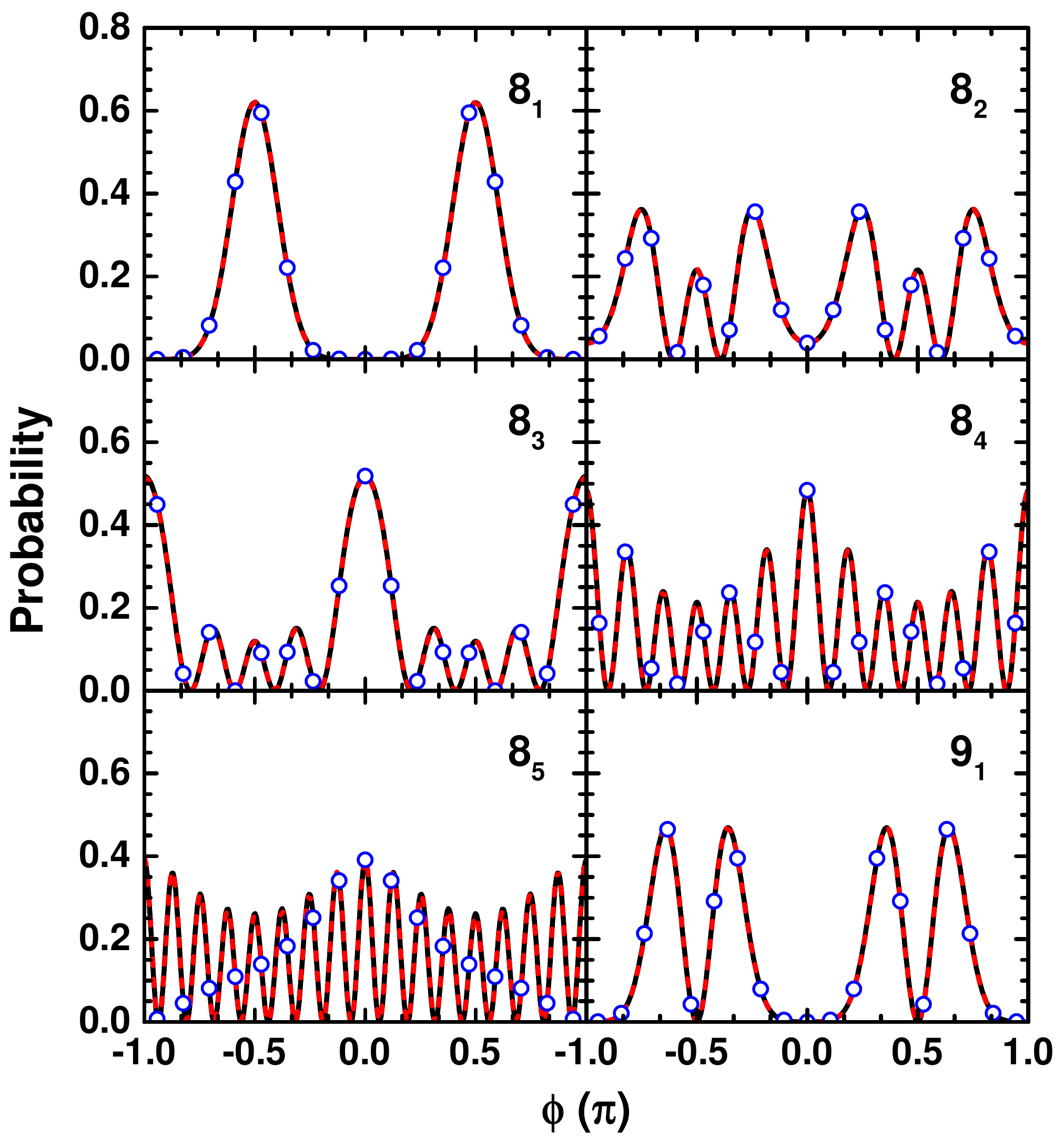}
    \caption{Probability densities of the SSS for the triaxial
    rotor states 1-5 of $I=8$ and the state 1 of $I=9$. The smooth solid and
    dashed lines continuous SSS $P(\phi)_{\nu}$ obtained by the TRM
    and collective Hamiltonian, respectively. The dots represent the
    discrete SSS $\frac{2\pi}{2I+1}P(\phi_n)_{\nu}$. }\label{f:SSS_TR}
  \end{center}
\end{figure}

\begin{figure*}[ht]
\begin{center}
    \includegraphics[width=16.5 cm]{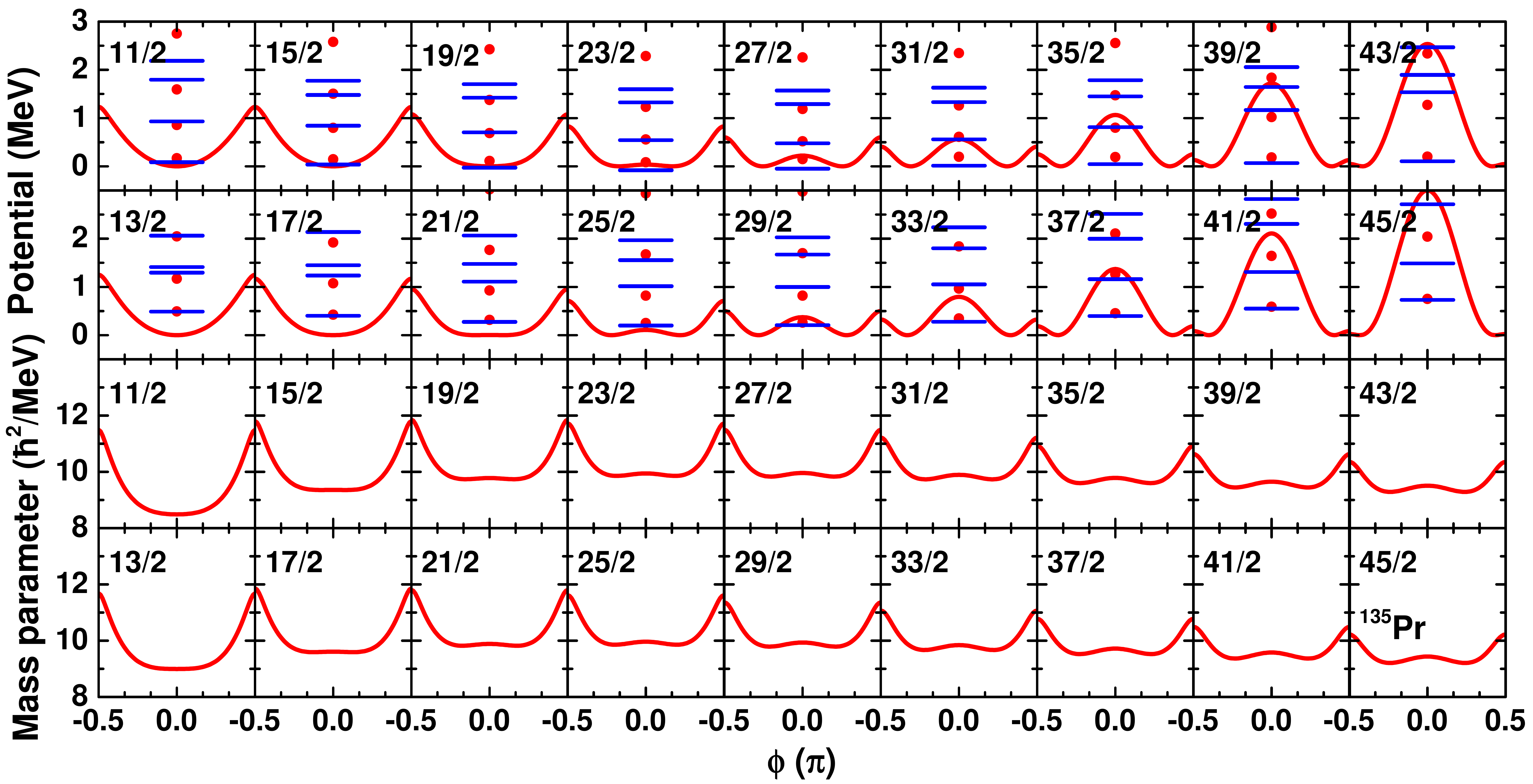}
    \caption{\label{f:potential_mass_135Pr} Collective potential energy
     $V(\phi)$ (upper two panels) and mass parameter $B(\phi)$ (lower two panels)
     as functions of $\phi$ for the particle-rotor system of $^{135}$Pr.
     The horizontal lines show the PTR energies $E_{\textrm{PTR}}$,
     while the dots show the collective Hamiltonian energies
     $E_{\textrm{CH}}$. Their energies are plotted with respect
     to the minima of the collective potential $E_{\textrm{ad}}(\textrm{min})$.}
\end{center}
\end{figure*}

Figure~\ref{f:Energy_TR} further displays the TRM energies for states with
angular momentum $I=8$ and $I=9$ as bars, while the energies obtained from
the collective Hamiltonian are represented by dots. It is important to note
that the $\textrm{D}_2$ symmetry imposes a restriction on the number of eigenstates.
Specifically, for $I=8$, there are five eigenstates labeled as $n=0$, $2$,
$4$, $6$, and $8$, while for $I=9$, there are four eigenstates labeled as
$n=1$, $3$, $5$, and $7$. Remarkably, the energies obtained from the collective
Hamiltonian exhibit perfect agreement with those derived from the TRM calculations.
This consistent agreement is expected because the derivation of the
collective Hamiltonian, given by Eq.~(\ref{eq2}), does not introduce
any approximations specifically for the triaxial rotor system. Consequently,
the excellent agreement confirms the validity and accuracy of the collective
Hamiltonian approach.

Interestingly, it is observed that the function $B(\phi)$ describing the
collective potential does not depend on the spin, as shown in Eq.~(\ref{BTRM}).
However, the term $V(\phi)$ in the potential does depend on the spin through
the factor $I(I+1)$, as described by Eq.~(\ref{VTRM}). As a result, the
stiffness of the collective potential increases with increasing spin. This
characteristic has significant implications, as it suggests that the
wobbling energy, defined as the energy difference between adjacent wobbling
states, also increases with spin. This finding provides a new perspective on
the conclusion that obtained by the assumption of small amplitude oscillations
of the total angular momentum in the framework of TRM in Ref.~\cite{Bohr1975}.

As aforementioned, the SSS plots offer a quantum mechanical perspective by
providing the probability density distribution of the wave function in the $\phi$
degree of freedom. These plots allow for an intuitive interpretation as they resemble
familiar probability density distributions in the $\phi$-representation. Hence,
the probabilities of the SSS are illustrates Fig.~\ref{f:SSS_TR} for the triaxial rotor
states numbered 1 to 5 with $I=8$, as well as for state 1 with $I=9$. These probabilities
are calculated using the collective Hamiltonian and are compared with the corresponding
probabilities obtained from the TRM calculations. Impressively, the resulting
SSS plots obtained from the collective Hamiltonian calculations exhibit a
high degree of agreement with those obtained from the TRM calculations.
This agreement suggests that the SSS plots derived from the collective Hamiltonian
can be interpreted as probability densities of the continuous eigen-functions
in the $\phi$-representation. Therefore, they provide a familiar and intuitive
interpretation of the system's quantum mechanical properties.

In the classical regime, the motion of the triaxial rotor is confined to the
regions between the potential energy function $V(\phi)$ and the bars, as
depicted in Fig.~\ref{f:SSS_TR}. In the SSS plot, the $n=0$ state denoted
as $8_1$ exhibits a prominent bump, indicating its zero-phonon character.
This bump represents the presence of zero-point oscillations, which arise
from quantum mechanical fluctuations even at the system's ground state.
On the other hand, the $n=1$ state labeled as $9_1$ possesses a one-phonon
character. In this case, the $\phi$-oscillation has a larger amplitude,
evident in the SSS plot where it shows a zero crossing at $\phi=\pm\pi/2$.
This observation corresponds to the behavior of the Hermite polynomial
$H_1$, which characterizes the wave function associated with one-phonon
excitations.

The $n=2$ state, denoted as $8_2$, exhibits the characteristics of a
distorted two-phonon state within the triaxial rotor system. In this
state, the amplitude of the oscillation in the $\phi$ degree of freedom
reaches an almost $\pi$ value. Examining the SSS plots associated with
the $8_2$ state, it is observed that within the allowed regions of motion,
there are two zeros symmetrically positioned at $\pm\pi/2$. This pattern
reflects the influence of the Hermite polynomial $H_2$ in describing
the wave function associated with this state.

The SSS plots corresponding to the states $8_3$ ($n=4$), $8_4$ ($n=6$),
and $8_5$ ($n=8$) exhibit unique characteristics in the triaxial rotor
system. These plots reveal standing wave patterns with a periodicity of
$2n$, where the potential energy function modulates the heights of the peaks.
In particular, the $n=8$ state $8_5$ primarily represents an almost pure
$K_l=8$ structure, with a relatively weak modulation observed in the SSS plot.
This modulation effect arises due to the interplay between the underlying
potential energy landscape and the quantum mechanical properties associated
with the state. For the $n=4$ state $8_3$, the SSS plot exhibits prominently
enhanced peaks at $\phi=0$ and $\pi$. This enhancement indicates the proximity
of the classical separatrix orbit concerning the wobbling motion with respect
to the $m$ axis and $l$ axis. It represents the characteristic \textit{flip}
mode (FW)~\cite{Q.B.Chen2022EPJA} that involves a transition between the two
orientations of the unstable $s$ axis, coupled with the medium moment of inertia.

Figure~\ref{f:SSS_TR} also includes the probability distribution of the discrete
$\phi_n$ states, which is given by $\frac{2\pi}{2I+1}P(\phi_n)_{\nu}$
with a prefactor in order to properly normalize it. The SSS states smoothly interpolate
between the discrete points of the $\phi_n$-distribution, resulting in a continuous
wave function with a density distribution. However, the discreteness of the
$\phi_n$ distribution tends to obscure the intuitive understanding of SSS
structures, especially as the number of peaks increases. Particularly for
the state $8_5$, the discrete distribution appears counter intuitive.
Since $K=8$ is almost a good quantum number, one would expect the probabilities
for all $\phi_n$ to be approximately equal. Contrary to this expectation,
a decrease towards $\phi=\pm \pi$ is observed in the discrete $\phi_n$ distribution.
This decrease is caused by the symmetrization of the state, specifically
the superposition $\left(\vert IIK\rangle+\vert II-K\rangle \right)/\sqrt{2}$.
This symmetrization leads to an interference between the $\phi_{\pm n}$ states.
The continuous SSS density distribution, on the other hand, behaves as anticipated.
It oscillates proportionally to $\cos^2(K \phi)$, while the envelope remains
roughly constant. This behavior arises from the constructive and destructive
interference patterns that originate from the coherent superposition of different
$\phi_n$ states. Therefore, the SSS plots highlights the interplay between continuous
and discrete aspects of the quantum system and offers valuable insights into the
symmetries and interference effects at play in the triaxial rotor system.

\subsection{Particle-rotor system}

For the particle-rotor system, we discuss the $^{135}$Pr reported in
Refs.~\cite{Matta2015PRL, Sensharma2019PLB} as the first example for
transverse wobbling of triaxial nuclei with normal deformation.
In Ref.~\cite{Frauendorf2014PRC}, the concepts of TW and LW was
proposed in the framework of PTR using the tentative experimental result
of $^{135}$Pr as example. In Ref.~\cite{Q.B.Chen2016PRC_v1}, the reported
wobbling bands in $^{135}$Pr were investigated using the collective
Hamiltonian based on the TAC approach. The experimental energy spectra
of both yrast and wobbling bands are well reproduced by the collective
Hamiltonian. But the total angular momentum is not a good
quantum number due to the rotational symmetry breaking in TAC.
In Ref.~\cite{Streck2018PRC}, the behavior of the collective rotor in
wobbling motion is investigated in PTR in the rotor angular momentum
representation. In Ref.~\cite{Q.B.Chen2022EPJA}, the interpretation
of the quantum states of the model of a triaxial rotor coupled to an
odd-particle was discussed in detail using the wobbling motion in
$^{135}$Pr as example. In particular, two-dimensional plots of the
probability distributions of the SCS were used to generalize the
classification of the collective excitations of the PTR as TW and LW
modes based on their topologies. In this subsection, we will show
the results obtained by the collective Hamiltonian and the
one-dimensional SSS.

The input parameters are $\beta=0.17$ (corresponds to $\kappa=0.038$),
$\gamma=-26^\circ$, and $\mathcal{J}_{m,s,l}=21$, 13, $4~\hbar^2/\textrm{MeV}$,
the same as in Refs.~\cite{Frauendorf2014PRC, Streck2018PRC, Q.B.Chen2022EPJA}.

To construct a collective Hamiltonian, similar as the triaxial rotor case,
we first calculate the collective potential and mass parameter staring from
the adiabatic particle-rotor Hamiltonian (\ref{eq7}). The adiabatic energy
$E_{\textrm{ad}}(\phi)$, which can be considered as the potential energy
for the collective wobbling mode, is obtained by minimizing adiabatic energy
$E_{\textrm{ad}} (\theta, \phi)$ (\ref{eq6}) with respect to the
angles $\theta$ for fixed $\phi$. The adiabatic energy
$E_{\textrm{ad}}(\phi)-E_{\textrm{ad}}(\textrm{min})$ is displayed
in Fig.~\ref{f:potential_mass_135Pr}, which represents the bottom of the valley
in the surface $E_{\textrm{ad}}(\theta, \phi)$ relative
to its lowest point. To simplify the presentation, we focus on
displaying the results within the region of $-\pi/2 \leq \phi \leq \pi/2$.
However, it is important to note that the results in the regions
of $-\pi \leq \phi \leq -\pi/2$ and $\pi/2 \leq \phi \leq \pi$
have symmetrical behavior with respect to the results in the region
$-\pi/2 \leq \phi \leq 0$ and $0 \leq \phi \leq \pi/2$, respectively.
This symmetry arises due to that the particle-rotor Hamiltonian is
invariant under the $\textrm{D}_2$ symmetry.

As shown in Fig.~\ref{f:potential_mass_135Pr}, with increasing spin $I$ the
adiabatic energy becomes biased toward the $m$ axis. Above the critical
angular momentum $J_c=11$ a maximum at $\phi=0$ appears, which
generates two minima with the same energy located at the angles
$\pm\phi$.

\begin{figure}[ht]
\begin{center}
 \includegraphics[width=\linewidth]{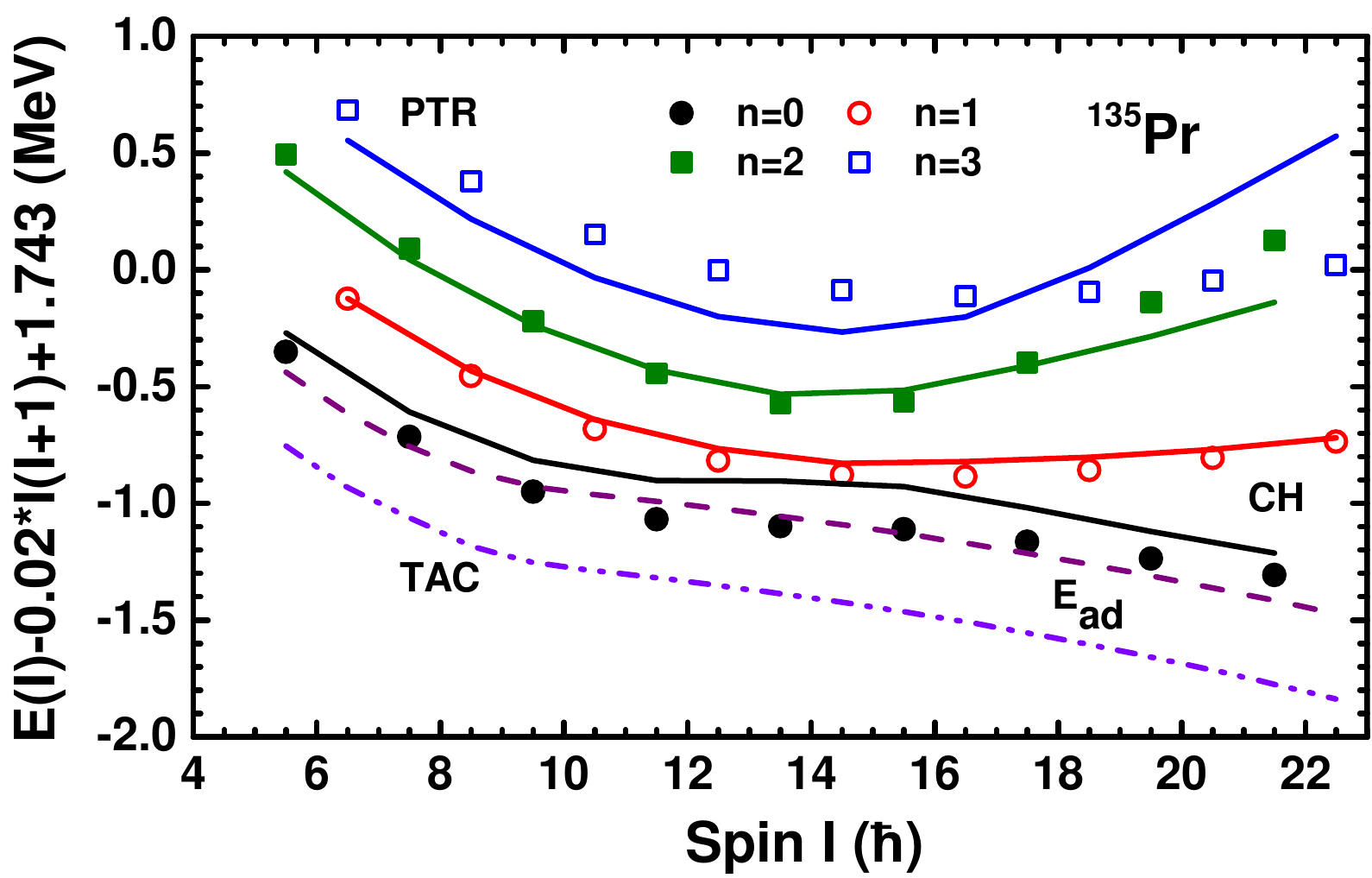}
 \caption{Energies of the lowest states $E_\nu$ of the
 PTR Hamiltonian (symbols) and the collective  Hamiltonian (full drawn lines),
 the adiabatic energy (dahed) and the TAC energy (dashed-dotted), for $^{135}$Pr.
 The energies are shifted by 1.743 MeV, which is the lowest energy
 from the diagonalization of $h_p(\gamma)$ in Eq.~(\ref{eq:hproton}).
 In the following, the yrast states ($n=0$) are denoted by
 $11/2_1$, $15/2_1$, $19/2_1$, ..., the single wobbling
 excitations ($n=1$) by $13/2_1$, $17/2_1$, $21/2_1$, ...,
 and the double wobbling excitations ($n=2$) by $15/2_2$, $19/2_2$,
 $23/2_2$, .... }\label{f:Energy_135Pr}
\end{center}
\end{figure}

\begin{figure}[ht]
\begin{center}
 \includegraphics[width=7.5 cm]{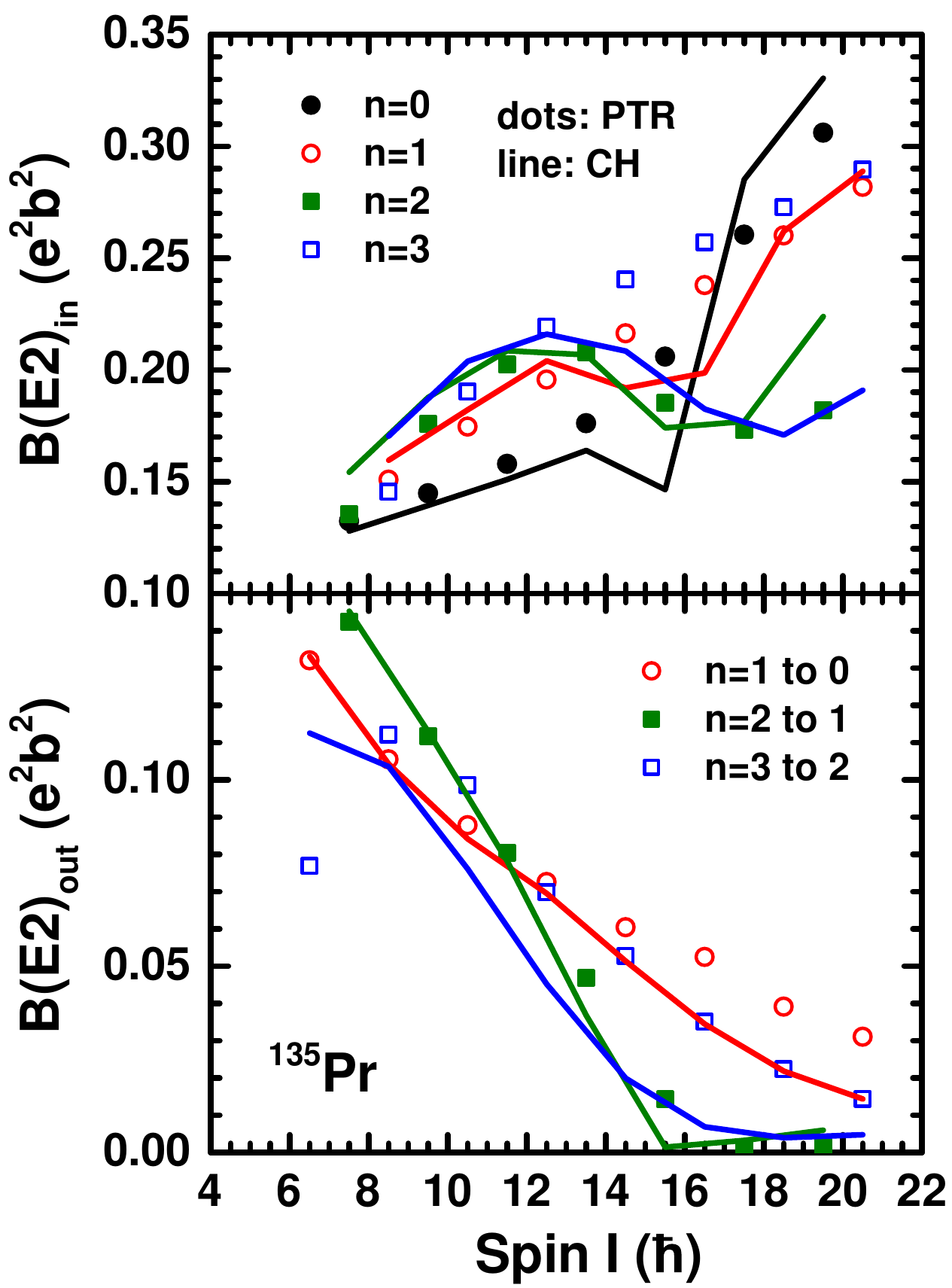}
 \caption{Upper panel: calculated in-band $B(E2)_{\textrm{in}}(I \to I-2)$, of
 the bands in Fig.~\ref{f:Energy_135Pr}, where the same line and color conventions
 are used. Lower panel: calculated inter-band $B(E2)_{\textrm{out}}(I \to I-1)$
 transition probabilities between the bands shown in Fig.~\ref{f:Energy_135Pr}.
 In order to keep the legend compact, the connecting transitions are labelled
 by quanta numbers $n$, which are associated with the states in the TW
 regime. Beyond, $n$ represents just a counting label for the states.}\label{f:BE2_135Pr}
\end{center}
\end{figure}

The mass parameters $B(\phi)$ calculated by Eq.~(\ref{eq12}) are shown
in Fig.~\ref{f:potential_mass_135Pr} as well. As expected for a triaxial nucleus with $\textrm{D}_{2}$
symmetry, $B(\phi)$ is symmetric with respect to $\phi=0$. The peaks of
$B(\phi)$ at $\phi=\pm \pi/2$  drive the nucleus
towards the $m$ axis with the largest moment of inertia. Unlike in the case of the
triaxial rotor system, the $B(\phi)$ in PTR changes with spin. The magnitude
of $B(\phi)$ increases with spin in the low spin region $I\leq 29/2$, while
decreases above $I=31/2$.

The adiabatic potential allows one to associate a collective wave function
with the wobbling mode, which provides a quantal perspective like
the TRM wave function in Sec.~\ref{s:TR}. The collective Hamiltonian (CH)
\begin{equation}\label{eq:HCH}
H_{\textrm{CH}}=\frac{J_3^2}{2B(\phi)}+V(\phi)
\end{equation}
is constructed by adding a kinetic term with
the inertia parameter $B(\phi)$ to the potential energy
$V(\phi)$ shown in Fig.~\ref{f:potential_mass_135Pr}. The CH
is diagonalized in the discrete basis (\ref{TRM-basis}) with $K$ being $I$,
$I-2$, ..., $-I+1$.
The pertaining amplitudes $C^{(\nu)}_{IK}$ represent the collective wave
functions. For half-integer spin there are two degenerate states, because
the CH couples  only $K$ with $K, K\pm 2$. They are
\begin{align}
 \vert \alpha_3\rangle &=\sum_K C^{(\nu)}_{IK} |IIK\rangle,\quad
 \vert -\alpha_3\rangle=\sum_K C^{(\nu)}_{IK} |II-K\rangle, \nonumber\\
 K&=I,~I-2,~...,~-I+1,
\end{align}
where $\alpha_3=I$ is the signature with respect to ${\cal R}_3(\pi)$.
The two solutions are related by time reversal and represent examples of Kramer's
degeneracy of half-integer spin systems. It is sufficient to study the
$\vert \alpha_3\rangle $.

The diagonalization of the CH provides more solutions than the PTR, because
the latter takes into account that the TR rotor core states are the completely
symmetric representations of the D$_2$ group. This entangles the quantum state
of the particle with the rotational state. The symmetry restriction gets lost
in deriving the CH. In the quantal PTR version of Eq.~(\ref{eq7}) the Coriolis
matrix elements $2\hat{J}_i\hat{j}_i$ couple the components $|IIK\rangle|jk\rangle$
in such a way that the change of the particle $j$ projection $k$ and the total $J$
projection $K$ are related such the core is total symmetric with respect
to D$_2$. This correlation gets lost when $\hat{J}_i$ is replaced by
the number ${J}_i$ in calculating the adiabatic energy surface (\ref{eq7}).

Comparing with the PTR solutions, one finds that for the signature $I=j+2m$ ($m$ being
an integer number) the first and third solutions describe the $n=0$ and $n=2$ wobbling
states. The second and fourth solutions should be discarded. These states with the $n=1$
and $n=3$ character do not appear in the PTR calculations. Likewise, for the signature
$I=j+1+2m$ the second and fourth solutions describe the $n=1$ and $n=3$ wobbling states.
The first and second solutions should be discarded. These states with the $n=0$ and $n=2$
character do not appear in the PTR calculations.

The resulting energies of the lowest bands, which represent the $n=0$, 1, 2, 3 wobbling
states, agree rather well with the PTR values as displayed in Figs.~\ref{f:potential_mass_135Pr}
and \ref{f:Energy_135Pr}. Only the $n=0$ yrast states are 0.1-0.2 MeV too high.
The horizontal lines in Fig.~\ref{f:potential_mass_135Pr} show the energies with
respect to the potential's minimum,
\begin{align}\label{eq:Ewa}
 E_w=E_\nu-E_{\textrm{ad}}(\textrm{min}),
\end{align}
where $E_\nu$ are the energies calculated by the exact diagonalization of
the PTR Hamiltonian or the CH  and $E_{\textrm{ad}}(\textrm{min})$, which
are shown as the dashed line in Fig.~\ref{f:Energy_135Pr}. The energy
difference (\ref{eq:Ewa}) can be assigned to the collective wobbling energy.

\begin{figure*}[ht]
\begin{center}
    \includegraphics[width=16.5 cm]{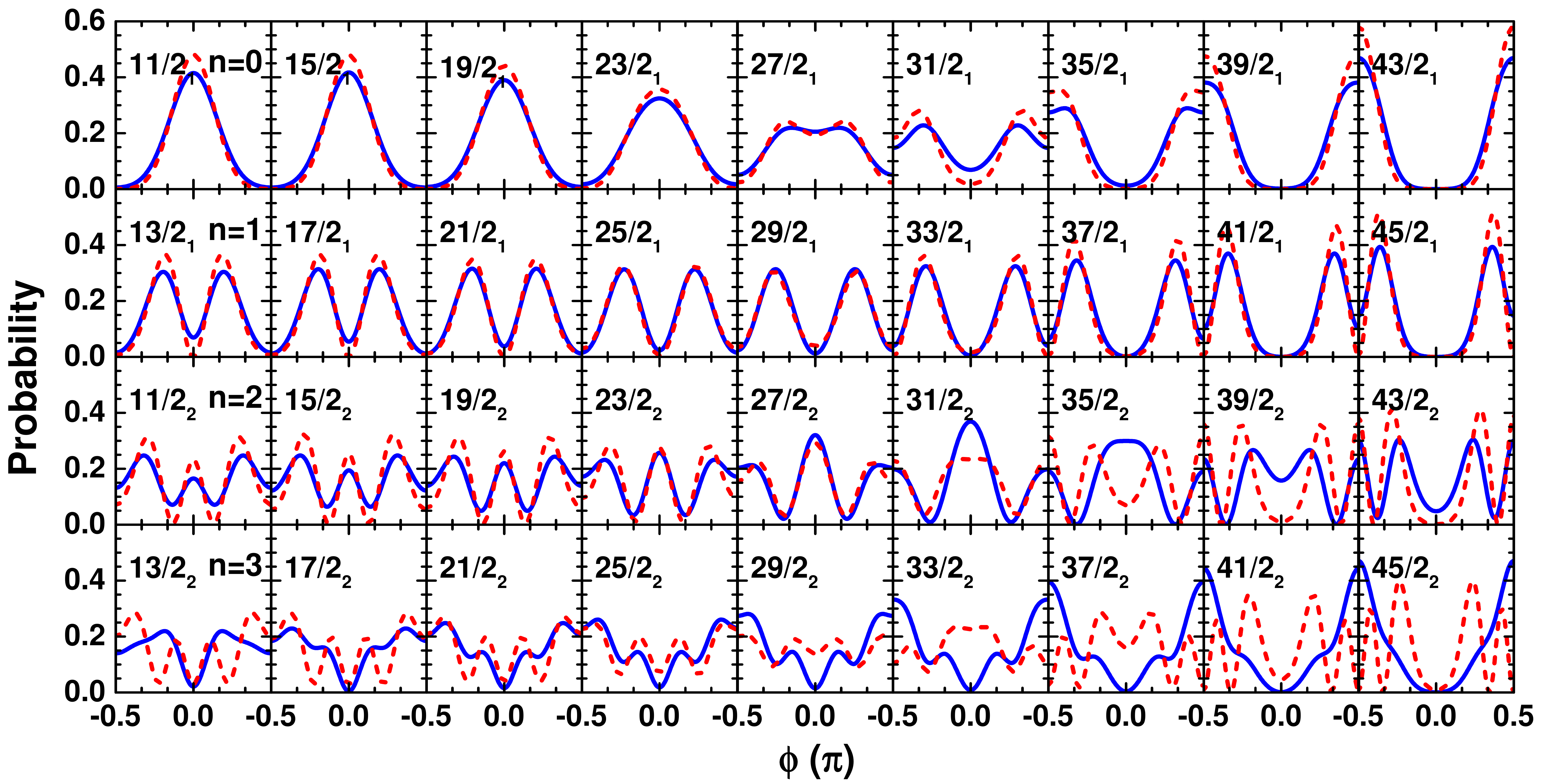}
    \caption{\label{f:SSS_135Pr} Full curves show the probability density
    of the SSS for the $n=0$-$3$ states calculated by PTR, while the
    dashed curves display the ones by the CH.}
\end{center}
\end{figure*}

\begin{figure*}[ht]
\begin{center}
    \includegraphics[width=15.0 cm]{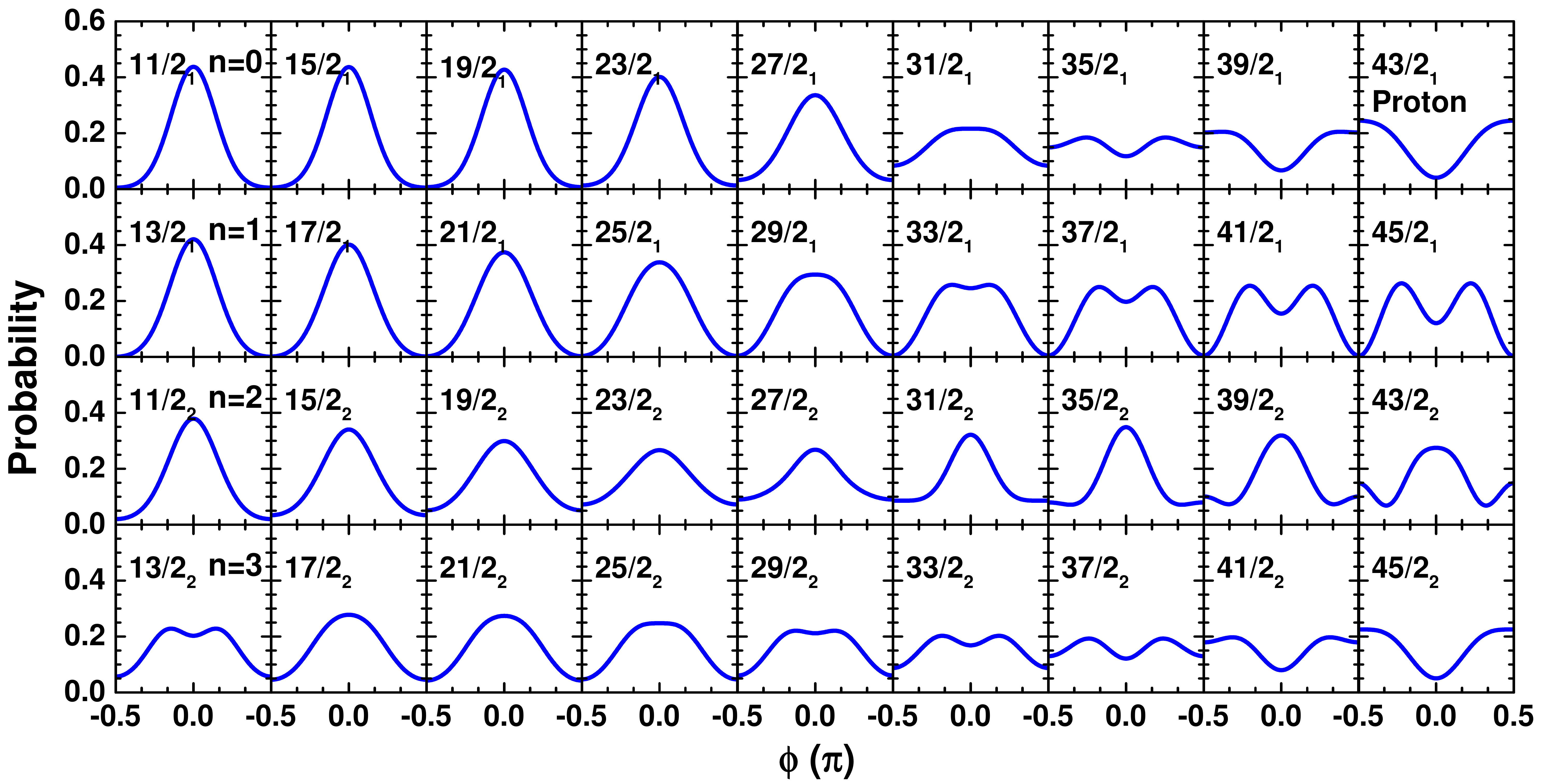}
    \caption{\label{f:SSS_j1} The SSS probability density for the
    odd particle calculated by PTR for the $n=0$-$3$ states.}
\end{center}
\end{figure*}

The upper panel of Fig.~\ref{f:BE2_135Pr} compares the CH intra-band
$B(E2)_{\textrm{in}} (I\to I-2)$ values with the PTR ones, where the
same line convention as in Fig.~\ref{f:Energy_135Pr} is used.
The lower panel shows the inter-band $B(E2)_{\textrm{out}} (I \to I- 1)$
transition probabilities between the $n=0$, 1, 2, and 3 bands in Fig.~\ref{f:Energy_135Pr}.
The strong collective $\Delta I=1$ $E2$ transitions, which characterize
the wobbling motion, are reflected by the Fig.~\ref{f:BE2_135Pr}.
The CH results become inaccurate around $I=33/2$, indicating that
the adiabatic approximation is not good anymore.
For $n=3$ band the adiabatic approximation is also not good enough.

The SSS collective (SSS-CH) probability densities are calculated as
the square of the collective wave function Eq.~(\ref{eq8}) obtained
from the pure density matrices of the $\alpha_3$ solutions. Using the
$-\alpha_3$ solutions gives the same results.

Figure~\ref{f:SSS_135Pr} compares these SSS-CH probability densities
with the SSS PTR (SSS-PTR) densities directly obtained from the PTR reduced
density matrices. Like for the TRM (see Sec.~\ref{s:TR}), we also diagonalized
the collective Hamiltonian within a sufficiently extended basis of
$\vert K\vert>I$-states. The results agree with ones shown in
Figs.~\ref{f:potential_mass_135Pr}-\ref{f:SSS_135Pr}
on the scale of the figures. Hence one can see the SSS-CH densities
as the probability densities of collective wave functions
in the continuous $\phi$ approximation.

For the states $n=0$ and $n=1$, SSS-PTR probabilities are rather similar
to the SSS-CH densities. This demonstrates that the development of the
structure is well accounted by the collective Hamiltonian composed of
the adiabatic double-well periodic potential and a kinetic term with a
$\phi$-dependent inertia parameter. The analog Hamiltonian describes
the relative ``pseudo-rotation" of the two parts of a molecule attached
to each other by one chemical bond (see, e.g., Ref.~\cite{Lewis1972JMS}), where
the collective wave functions are characteristic for double-well periodic
potentials. The probability distributions in $\phi$ direction are standing waves with
1, 2, 3, 4 maxima in the interval $-\pi/2 \leq \phi \leq \pi/2$ for $n=0,~1,~2,~3$,
respectively. Within the full range $-\pi \leq \phi \leq \pi$ the maxima are
located symmetrically to $\phi = 0,~\pm\pi/2,~\pm\pi$.
Depending on the position of the potential minimum, the maxima form pairs
close  to $\phi=0, ~\pm\pi$ in the case of TW and $\phi=\pm \pi/2$
in the case of LW.

The SSS-PTR densities correlate well with the SSS-CH ones of the $n=0$
and 1 collective states. The $n=2$ double wobbling structure is strongly
disturbed, yet recognisable. For $n=3$ the SSS density substantially deviates
from the CH one. The adiabatic approximation,
upon which the concept of a collective wave function is based,
becomes progressively inaccurate with increasing $n$.

There is a complementary perspective. In calculating the reduced density matrix,
the tracing out of the $\bm{j}$-degree of freedom destroys to some extend the
coherence of the complete PTR wave function. In general, the properties of
the $\bm{J}$-degrees of freedom are described by the reduced density matrix,
which cannot be further simplified. If $\bm{j}$ follows $\bm{J}$
in an adiabatic way, the reduced density matrix can be approximated
by a pure density matrix generated from a collective wave function
in the $\bm{J}$-degree of freedom. The coherence in the $\bm{J}$-degree of
freedom decreases with increasing $n$. One can quantify the degree of coherence
by diagonalizing the reduced density matrix. For a pure, completely coherent state
one eigenvalue is 1 and all other are zero. The TRM states are examples.
For partial coherence one has one large eigenvalue, the eigenvector of which
represents the coherent wave function and the eigenvalue its probability.
The remaining eigenfunctions, which appear with the probability of their
small eigenvalues, distort the coherence.

\subsubsection{The TW regime}

For $I\leq 23/2$, the collective potential has a minimum at $\phi=0$.
The SSS-PTR probability distributions in Fig.~\ref{f:SSS_135Pr}
are similar to the SSS-CH of the collective wave functions with $n=0$,
1, and 2 within the adiabatic potential which is centered at $\phi=0$.
The densities have $n$ pronounced maxima symmetric to $\phi=0$, which
reflect the Hermite polynomials $H_n$. The behavior is analog to the
TRM discussed in the context of Figs.~\ref{f:Energy_TR} and \ref{f:SSS_TR}.
With increasing $n$, the SSS-PTR distributions appear progressively
washed out as compared to the SSS-CH distributions of the collective
wave functions. We attribute this to a loss of coherence which signals
that the adiabatic approximation loses accuracy. Figure~\ref{f:SSS_135Pr} shows
that the $n=3$ PTR structures are not well described  by the CH.

The SSS-PTR distribution of state $n=3$ substantially deviates from the
SSS-CH of the collective wave functions, which indicates that the adiabatic
approximation fails. This can be seen comparing the SSS distribution for
the odd particle in Fig.~\ref{f:SSS_j1}. As expected for the adiabatic behavior,
the particle is well aligned with the $s$ axis for $n=0$, $1$, and $2$.
The $n=3$ distribution is different. We associated the difference with the
mixing of the $n=3$ collective wobbling state and the signature partner band.
The latter represents the first excited state of the adiabatic Hamiltonian
(\ref{eq7}), which is not taken into account when deriving the CH.

For the $n=3$ states $17/2_2$, $21/2_2$, and $25/2_2$, the adiabatic condition
is as badly realized as for $13/2_2$. The SSS distributions for $\bm{j}$ are nearly
the same as for $13/2_2$ and $25/2_2$ in Fig.~\ref{f:SSS_j1}. The SSS-PTR densities
for $\bm{J}$ resemble the SSS-CH qualitatively. The classification of the states as
strongly perturbed three-phonon TW excitations seems acceptable. In our preceding
paper we argued that the coupling of the $n=3$ wobbling state with the signature
partner state becomes weaker with increasing $I$.

Up to $I=23/2$, the $n=0$ SSS distributions have a peak at $\phi=0$,
which gets wider with $I$. As expected for TW, the wobbling energy
\begin{align}\label{eq:Ew}
 E_w(I)=E(I,~n=1)-
 \frac{1}{2}\left[E(I-1,~n=0)+(I+1,~n=0)\right]
\end{align}
decreases with $I$.  The PTR transition probabilities in Fig.~\ref{f:BE2_135Pr}
are well reproduced by the CH states.


\subsubsection{The flip regime}

For $I=27/2$-$35/2$, the potential develops a maximum at $\phi=0$, which
generates two minima near $\phi=\pi/4$. The $n=0$ states $27/2_1$ and $31/2_1$ can be
interpreted as being composed of two states localized in the two minima (localized
states), which are coupled by tunneling through the shallow barriers. The superposition
of the localized states is constructive as seen by the substantial SSS probability
under the barriers. With increasing $I$, the tunneling through the barriers
at $\phi=0$ decreases and the tunneling through the barriers at
$\pm \pi/2$ increases, which reflects the respective change of the
barrier heights. For $I\geq 35/2$, the SSS plots are well understood in
terms of the even $n=0$ and odd $n=1$ wave functions in a collective potential
with an increasing barrier at $\phi=0$ and a decreasing small hump at
the $m$ axis. The $n=1$ states $29/2_1$ and $33/2_1$ represent the odd linear
composition of the local states. They are little sensitive to the presence of
the barriers at $\phi=\pm\pi/2$, because their probability is
small there, where the wave functions changes sign.

The SSS plots of the $n=2$ states can be understood in terms of the
probability densities of the collective wave functions belonging to the
periodic potentials shown in Fig.~\ref{f:potential_mass_135Pr}. The SSS plots
of the states $31/2_2$ and $35/2_2$ show two minima half-way between the
$s$ and $m$ axes and two maxima located at the barriers (at $\phi=0$,
$\pm\pi/2$). The SSS plot of the state $43/2_2$ shows the characteristic
two minima located symmetrically to $\phi=\pm\pi/2$, respectively.
The SSS probability of state $29/2_2$ illustrates the transition from
the TW to the flip mode (FW) and the one of the state of $39/2_2$ the
transition from the FW to the LW mode.

\begin{figure}[ht]
\begin{center}
    \includegraphics[width=7.0 cm]{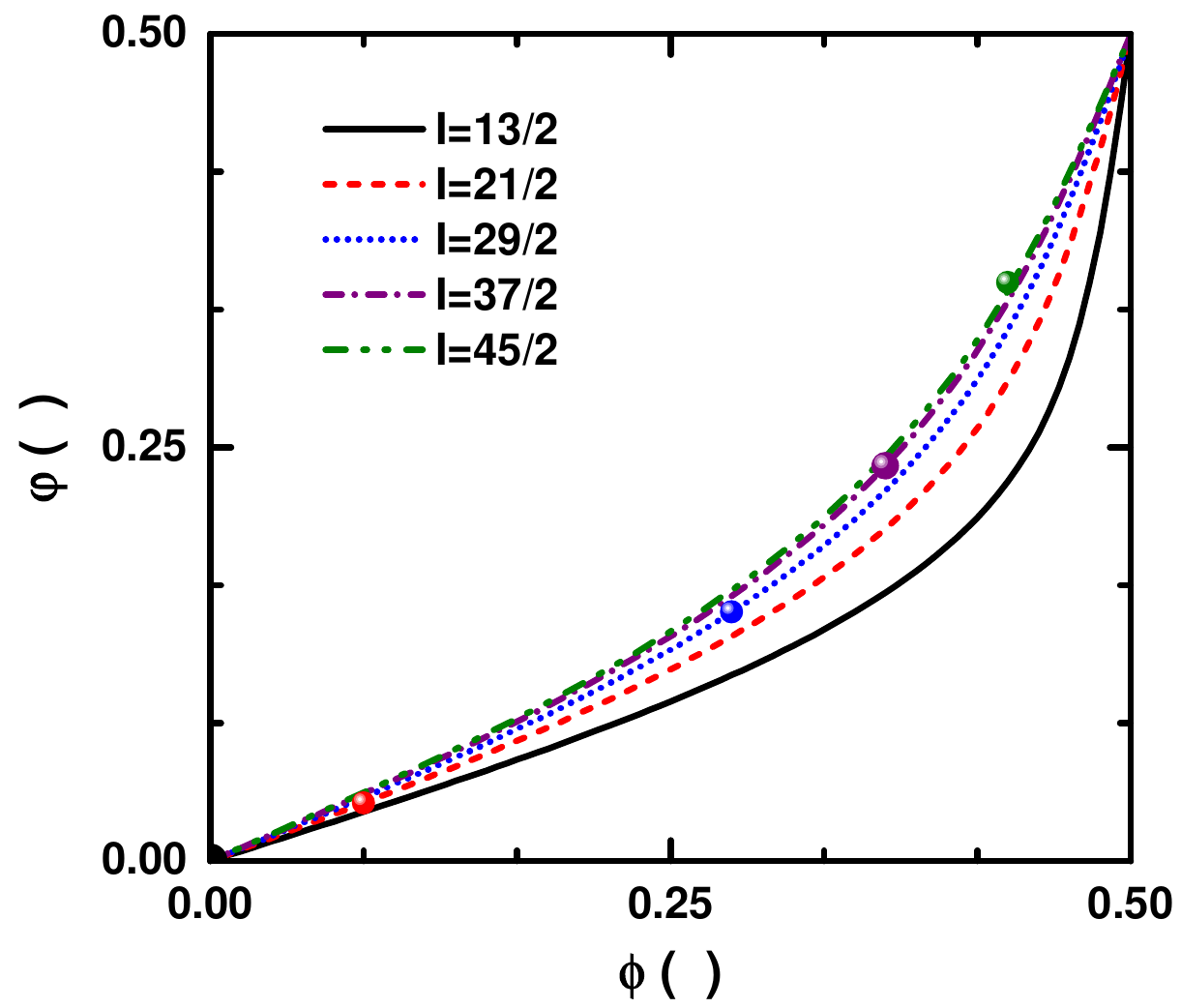}
    \caption{\label{f:phi_j} The azimuthal angle $\varphi$ of $\bm{j}$ as a function
    of azimuthal angle $\phi$ of $\bm{J}$ obtained by diagonalization of
    $H_{\textrm{ad}}$ for the selected spins $I=13/2$, $21/2$, $29/2$,
    $37/2$, and $45/2$. The dot on the curve label the potential minimum
    for each spin.}
\end{center}
\end{figure}

There is another perspective. The vicinity of the barrier top slows
down the motion, which is reflected by the large probability density.
Accordingly, the vector $\bm{J}$ flips between the barrier tops
like $8_3$ state of the TRM, which is close to the separatrix.

\begin{figure*}[ht]
\begin{center}
    \includegraphics[width=14.0 cm]{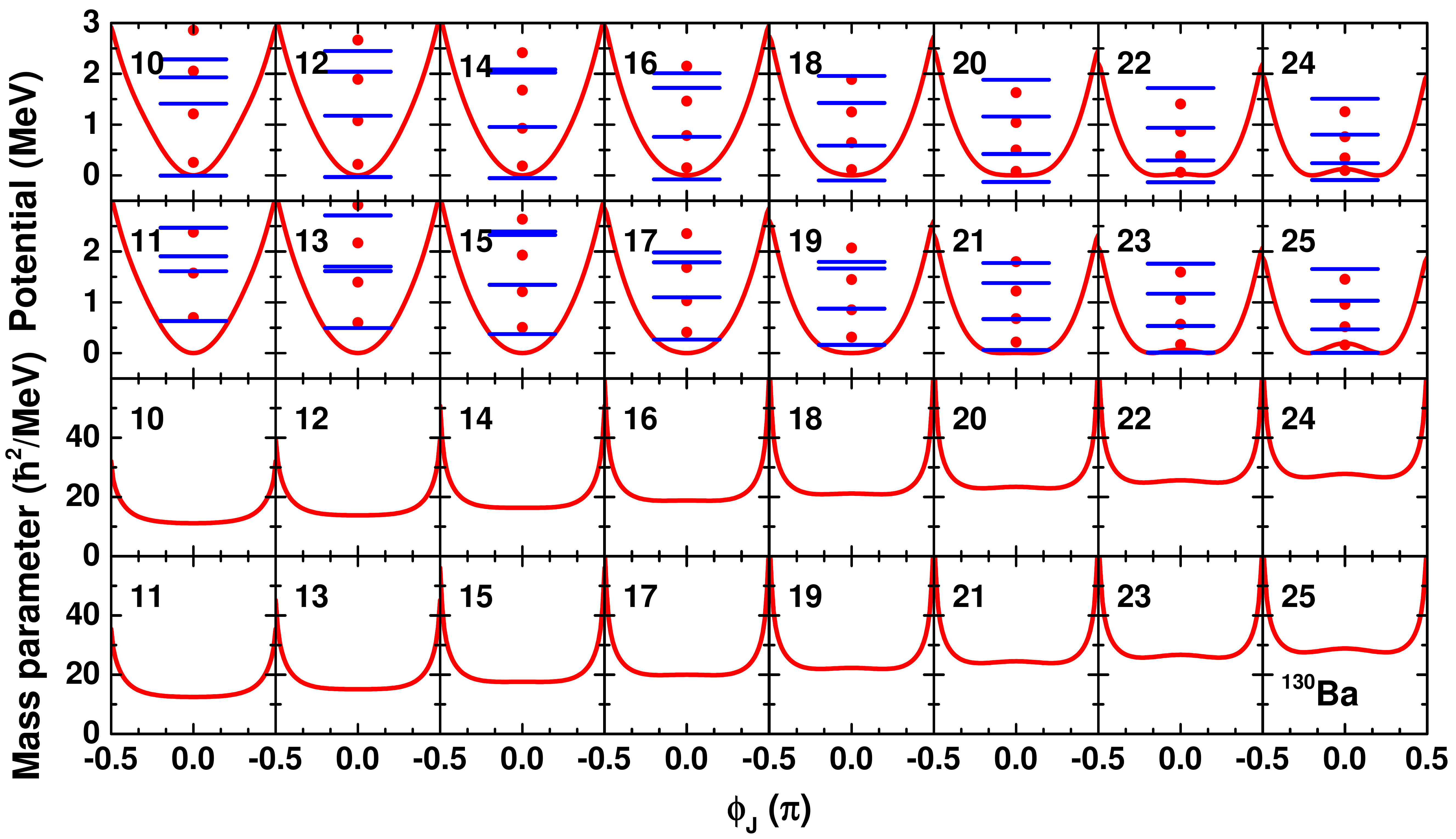}
    \caption{Same as Fig.~\ref{f:potential_mass_135Pr}, but for the case of
    $^{130}$Ba.} \label{f:potential_mass_130Ba}
\end{center}
\end{figure*}

\subsubsection{The LW regime}

For the states $37/2_1$, $39/2_1$, $41/2_1$, and $45/2_1$, the
adiabatic potential in Fig.~\ref{f:potential_mass_135Pr} has deep
minima centered around $\phi=\pm \pi/2$ with a tiny bump in the
middle. The SSS-CH distributions follow  well the SSS-PTR ones. The
states have the nature of $n=0$ and 1 LW wobbling vibrations around
$\phi=\pm\pi/2$. The LW character is not generated by the ``frozen
alignment" of the proton angular momentum $\bm{j}$ with the $m$
axis. Figure~\ref{f:phi_j} shows the adiabatic angle $\varphi$ of
the particle vector $\bm{j}$ as function of the angle $\phi$ of the
total angular momentum $\bm{J}$. In case of the high $I$ values of
the LW regime, $\bm{j}$ is pulled toward $\bm{J}$ by the Coriolis
force but $\varphi<\phi$ due to the non-rotating which favors
$\varphi=0$. This explains the SSS plots in Fig.~\ref{f:SSS_j1}.
They show the same kind of maxima as the SSS-PTR. However, the
corresponding maxima are wider and shifted toward $\varphi=0$. The
difference is understood by means of Fig.~\ref{f:phi_j}.

For the states $33/2_2$, $37/2_2$, $41/2_2$, and $45/2_2$, the SSS densities
differ qualitatively from each other. As expected, the SSS-CH develops into a $n=3$
LW structure with three minima located symmetrically to $\phi=\pm\pi/2$.
In contrast, the maxima of the SSS-PTR distribution at $\phi=\pm \pi/2$
become dominant and the oscillations disappear. This indicates a change of
the structure, which is beyond the realm of adiabaticity. As seen in
Fig.~\ref{f:SSS_j1}, the particle SSS has a maximum whereas the SSS-PTR has
a minimum, i.e., $\bm{j}$ and $\bm{J}$ move with opposite phase, which is
not uncommon for higher excited states of coupled oscillators.

\subsection{Two-quasiarticle-rotor system}

In the study of two-quasiparticle-rotor system, the even-even nucleus $^{130}$Ba has
been identified as the first example of a system exhibiting two-quasiparticle wobbling
bands~\cite{Petrache2019PLB, Q.B.Chen2019PRC_v1}. The Ref.~\cite{Petrache2019PLB} observed
two new bands built on the two-quasiparticle $\pi(1h_{11/2})^2$ configuration in
$^{130}$Ba. In Ref.~\cite{Q.B.Chen2019PRC_v1}, these two bands were investigated
using the constrained triaxial covariant density functional theory (CDFT)
combined with quantum PTR calculations. The energy difference between the two
bands, as well as the available electromagnetic transition probabilities
$B(M1)_{\textrm{out}}/B(E2)_{\textrm{in}}$ and $B(E2)_{\textrm{out}}/B(E2)_{\textrm{in}}$,
were well reproduced. The analysis of the angular momentum geometry demonstrated
that lower band, the well-known $S$-band, represents the zero-phonon wobbling
state and the higher band represents the one-phonon TW state generated by a
two-quasiparticle configuration. Later on, the two-quasiparticle wobbling in
$^{130}$Ba was further examined using the projected shell model~\cite{Y.K.Wang2020PLB},
providing additional support for this phenomenon. In the following, we
use this example to discuss its  interpretation in terms of the
SSS visualization and a collective Hamiltonian.

\begin{figure}[ht]
\begin{center}
    \includegraphics[width=\linewidth]{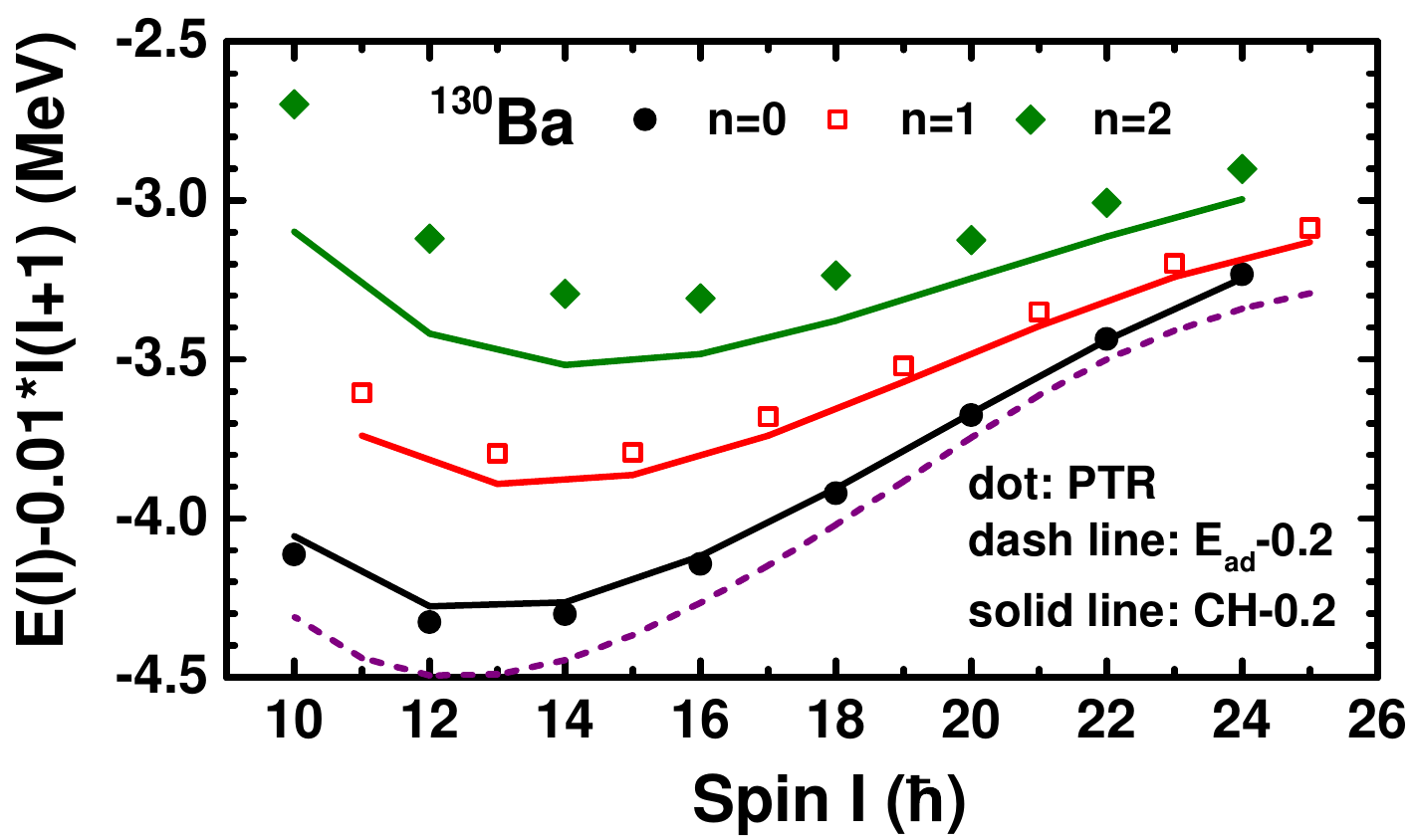}
    \caption{Same as Fig.~\ref{f:Energy_135Pr}, but for the case of
    $^{130}$Ba. Note that the energies obtained from the collective Hamiltonian
    are 0.2 MeV shifted down.}
    \label{f:Energy_130Ba}
\end{center}
\end{figure}

\begin{figure}[ht]
\begin{center}
 \includegraphics[width=7.0 cm]{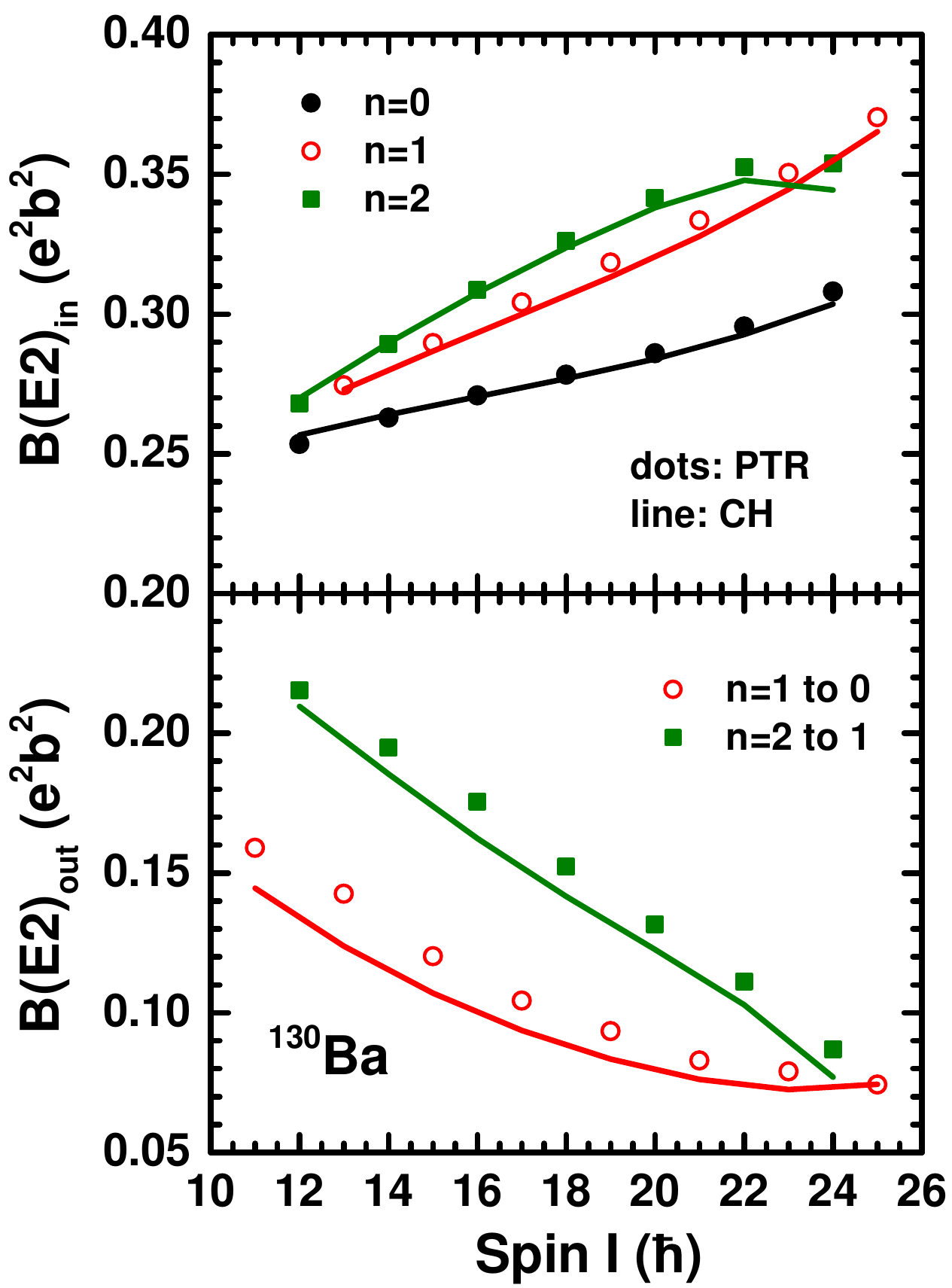}
 \caption{Same as Fig.~\ref{f:BE2_135Pr}, but for the case of $^{130}$Ba.}\label{f:BE2_130Ba}
\end{center}
\end{figure}

\begin{figure*}[ht]
\begin{center}
    \includegraphics[width=16.0 cm]{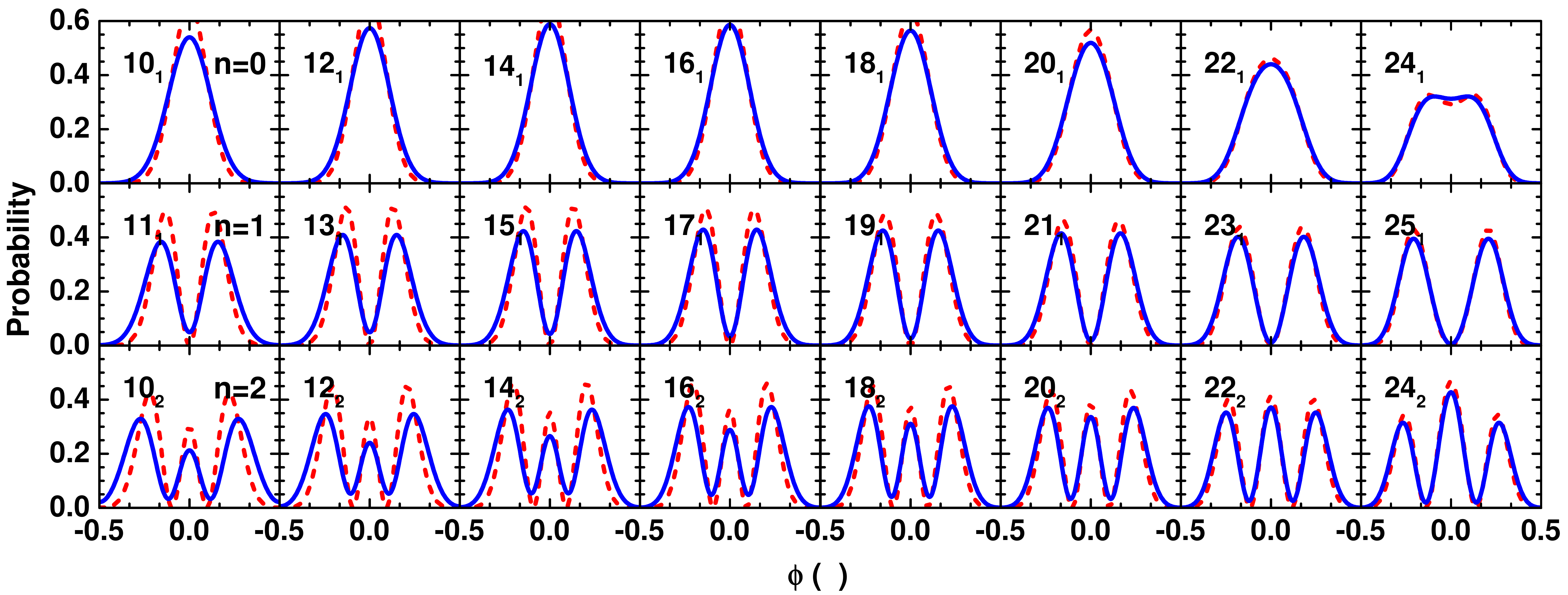}
    \caption{Same as Fig.~\ref{f:SSS_135Pr}, but for the case of $^{130}$Ba.}\label{f:SSS_130Ba}
\end{center}
\end{figure*}

In the investigation of the two-quasiparticle $\pi(1h_{11/2})^2$ configuration in
$^{130}$Ba~\cite{Q.B.Chen2019PRC_v1}, the deformation parameters were determined
to be at a triaxial shape characterized by ($\beta=0.24$, $\gamma=21.5^\circ$)
using constrained triaxial CDFT calculations. In order to model this triaxial
rotor system, three spin-dependent moments of inertia $\mathcal{J}_i=\Theta_i(1+cI)$
were introduced, where $i = s$, $m$, and $l$. The parameters $\Theta_{s,m,l}=1.09$,
1.50, and 0.65~$\hbar^2/\textrm{MeV}$ and $c=0.59$ were determined by adjusting
the PTR energies to the experimental energies of the zero- and one-phonon
bands~\cite{Q.B.Chen2019PRC_v1}.

The approach is analogous to the treatment of the one-particle case Eq.~(\ref{eq7}) ff.,
except the adiabatic Hamiltonian (\ref{eq7}) is diagonalized in the space of
the two-quasiparticle configurations. The adiabatic potential
$V(\phi)=E_{\textrm{ad}}(\phi)-E_{\textrm{ad}}(\textrm{min})$
and the mass parameter $B(\phi)$ are depicted in Fig.~\ref{f:potential_mass_130Ba}.
As the spin $I$ increases, the adiabatic energy becomes increasingly favored along the
$m$ axis, similar to the case of $^{135}$Pr shown in Fig.~\ref{f:potential_mass_135Pr}.
Beyond the critical angular momentum of $J_c=22$, a maximum at $\phi=0$ emerges,
resulting in the presence of two minima with equal energy at angles
$\pm\phi$. Compared to $^{135}$Pr, the transverse geometry in $^{130}$Ba
exhibits greater stability and is sustained over a much wider range
of spin values. The increased stability can be attributed to the
larger angular momentum aligned with the $s$ axis, which is generated
by two quasiprotons instead of one.

The mass parameter $B(\phi)$ is also presented in Fig.~\ref{f:potential_mass_130Ba}.
Its magnitude increases with increasing spin and is generally larger than the one
of $^{135}$Pr. This enhancement in the magnitude of $B(\phi)$ can be attributed
to the larger alignment along the $s$ axis caused by the presence of two protons.
The two protons contribute significantly to the overall dynamics of the system,
thus influencing the wobbling behavior.

The CH is diagonalized in the discrete basis (\ref{TRM-basis}) with even
$K$ being $I$, $I-2$, ..., $-I$ for the even-$I$ and $I-1$, $I-3$, ..., $-(I-1)$
for the odd-$I$. The resulting energies come in close pairs. For even-$I$
the first, third, ... solutions are associated with the $n=0$, 2, ... PTR wobbling states.
For odd-$I$ the third, fifth, ... solutions are associated with the $n=1$, 3, ... PTR
wobbling states. The selection can be justified by direct comparison with the PTR
solutions.

As the CH couples only $K$ with $K\pm 2$, there is a complementary set of solutions
obtained by diagonalizing the CH in the basis of odd-$K$. The energies turn out to
be very close to the energies of the even-$K$ solutions. The pertaining SSS
plots are indistinguishable from the ones of the even-$K$ ones on the scale
of Fig.~\ref{f:SSS_130Ba}. The full PTR state is composed of both even-
and odd-$K$ components. Thus the difference between the two types is a
consequence of approximation scheme. For this reason, the odd-$K$
solutions will be disregarded as well.

Figure~\ref{f:Energy_130Ba} compares the PTR energies with the adopted CH energies.
The agreement is better than in the case of $^{135}$Pr. Only the $n=2$ states are
somewhat too low. The CH potentials in Fig.~\ref{f:potential_mass_130Ba} indicate
that in $^{130}$Ba the TW regime extends to $I=22$ where the flip regime begins.
Accordingly, the energies change from the equidistant harmonic TW spectrum at
low $I$ to $E_w \approx 0$ at $I=24$, which indicates the instability of TW.
The adiabatic approximation very well accounts for the structural change.

In Fig.~\ref{f:BE2_130Ba}, the upper panel displays the in-band
$B(E2)_{\textrm{in}} (I\to I-2)$ transition probabilities for the bands
presented in Fig.~\ref{f:Energy_130Ba}, where the same line conventions
are employed. The lower panel exhibits the inter-band $B(E2)_{\textrm{out}} (I \to I- 1)$
transition probabilities for the $n=0$, 1, and 2 bands depicted
in Fig.~\ref{f:Energy_130Ba}. It is noted that the wobbling motion's
strong collective $\Delta I=1$ $E2$ transitions are clearly
evident in Fig.~\ref{f:BE2_130Ba}. The collective Hamiltonian
describes very well the $B(E2)_{\textrm{in}} (I\to I-2)$ PTR values.
The $B(E2)_{\textrm{out}} (I \to I- 1)$ of the collective Hamiltonian
are overall slightly smaller than those of PTR.

Figure~\ref{f:SSS_130Ba} compares the SSS-CH probability densities calculated
from the accepted collective wave functions with the SSS-PTR densities derived
from the PTR reduced density matrices. Clearly, the adiabatic approximation
works well for all cases. The distinctive features associated with the wobbling
states $n=0$, 1, and 2 manifest themselves as the increasing number of minima
symmetrically situated around $\phi=0$. They reflect the Hermit polynomials
which characterize the number of excitation quanta in the system.

The remarkable correspondence between the SSS-PTR densities and the SSS-CH
densities affirms that the structural evolution can be effectively described
by the collective Hamiltonian, which comprises the adiabatic double-well periodic
potential and a kinetic term incorporating a $\phi$-dependent inertia parameter.
Hence, it can be concluded that the implementation of SSS-plots in the analysis
of the one-quasiparticle-rotor and two-quasiparticle-rotor systems leads to
intuitive and comprehensive understanding of the underlying phenomena.


\subsection{Collective Hamiltonian based on Tilted Axis Cranking}

Figure~\ref{f:Energy_135Pr} also includes the energy of $n=0$ band calculated
by the TAC approach. The total Routhian
\begin{align}\label{ee:HTAC}
 \hat{H}_{\textrm{TAC}}
 &=\hat{h}_p-\omega(\cos \phi\hat{j}_1+\sin \phi\hat{j}_2)\notag\\
 &-\frac{1}{2}\mathcal{J}_1\omega^2\cos^2\phi-\frac{1}{2}\mathcal{J}_2\omega^2\sin^2\phi
\end{align}
is diagonalized for a grid of cranking frequency $\omega$ and tilted angle
$\phi$ values, which gives the Routhian $E^\prime_{\textrm{TAC}}(\omega,\phi)$
as the lowest eigenvalue and the pertaining classical angular vector components
\begin{align}
 J_1(\omega,\phi)=\langle \hat{j}_1\rangle+\omega \cos \phi \mathcal{J}_{1}\\
 J_2(\omega,\phi)=\langle \hat{j}_2\rangle+\omega \sin \phi \mathcal{J}_{2}.
\end{align}
Whereas for the above discussed adiabatic PTR the recoil term $\langle \hat{j}_i^2 \rangle$
is diagonalized, it is  replaced by $\langle \hat{j}_i\rangle^2$ under the
TAC mean field approximation. In TAC the total angular momentum $J$
is constrained to be $\sqrt{I(I+1)}$ by adjusting the cranking frequency $\omega$.
The difference between the TAC energy and $E_{\textrm{ad}}$ are almost
constant 0.3 MeV in the whole spin region. That is, up to this shift the classical
energy is the same as the PTR adiabatic energy, and the CH based on the TAC
classical energy is nearly the same as well.

The authors of Refs.~\cite{Q.B.Chen2014PRC, Q.B.Chen2016PRC_v1} calculated
the TAC energy and used the Routhian as the collective potential
$V(\phi)=E^\prime_{\textrm{TAC}}(\omega,\phi)$ for a fixed value of the
frequency $\omega$. The mass parameter was estimated using the
frozen alignment (FA) approximation~\cite{Frauendorf2014PRC}, namely a
harmonic vibration around $\phi=0$. Based on these, the CH method
gives wobbling energies that are in agreement with the PTR
values.

The success of the CH based on the TAC classical energy suggests applying
the method developed for the PTR system to the microscopic versions of the
TAC approach. The problem that the CH provides to many solutions will arise
as well. The ``exact" PTR solution which we used to select the appropriate
solution will not be available in this case. In a future study of the CH based
on the microscopic TAC
we will discuss how the selection can be based on the analysis of
the symmetry-broken states of the adiabatic Hamiltonian (\ref{eq7}).


\section{Summary}

In summary, we introduced the Spin Squeezed States (SSS) as a non-orthogonal and
over-complete basis. The utilization of SSS plots in the analysis of the particle-rotor
system offers a new perspective into the angular momentum geometry and complex dynamics
of the constituents by visualizing the system in terms of the familiar paradigm of
a potential and a mass parameter for the orientation of the total angular momentum.

The semiclassical approach was combined with the adiabatic approximation in order to
calculate the collective potential and mass parameters of a collective Hamiltonian (CH)
capable of describing the wobbling motion in both even-even and odd-mass systems.
The adiabatic approximation assumes that the response of the one or two particles
can be calculated by classically rotating the triaxial potential they are moving
therein, which, in essence, amounts to the tilted cranking method. By diagonalizing
the resulting CH, we obtained the energies, $E2$ transition probabilities,
and wave functions associated with the wobbling states, which were
compared with the exact ones from the diagonalization of the original one- and
two-particle plus triaxial rotor Hamiltonian. The SSS plots turned out to a
most useful for comparing the wave functions and generating a familiar
quantum mechanical perspective on wobbling motion.

By analyzing the SSS plots of the probability density as a function of the
angle $\phi$ around the long axis, we interpreted them as the probability
density of collective wave functions associated with the collective periodic
potential $V(\phi)$ (equal to the adiabatic classical energy). By fixing the
energy to one of the PTR states, the classical motion is confined to stay
within $V(\phi)$, which provides a straightforward topological classification.
For the TW mode the allowed region is centered around one of the axes
perpendicular to the medium axis with the largest moment of inertia. For
the LW mode the allowed region is centered around the medium axis itself.
For the considered cases of one or two particles at the bottom of a
high-$j$ shell, with increasing angular momentum $V(\phi)$ changes from
being centered around the short axis ($\phi=0, \pm \pi$) to
being centered around the medium axis ($\phi=\pm \pi/2$). That is,
the mode changes from TW to LW. In the transition region the potential minima
are located around $\phi=\pm \pi/4, ~\pm 3\pi/4$. The lowest states
represent FW modes where the quantum system jumps between different regions
of the potential landscape. The $n=0$ state flips between the minima with
no phase change. The $n=1$ state flips between the minima with a phase change
the at $\phi=0,~\pi$. However, for the $n=2$ state
when the energy of the system is close to the tops of the potential
barriers, the system flips between these locations.

For our example of $^{135}$Pr, the SSS plots calculated by the exact
PTR model correlate very well with the probability densities of
the wave functions that belong to the CH. One recognizes
the TW, LW, and transitional FW pattern implied by
the collective Hamiltonian for $n=0$, $1$ wobbling states. For
the $n=2$ state the exact density profile shows the triple maxima
of the collective wave function in the TW and LW regime. However,
it deviates from the flip profile between the potential barriers.
The exact profile of the next excitation differs qualitatively
from the $n=3$ structure of the CH. The increasing deviations with
$n$ of the SSS plots of the collective wave functions from the
exact ones reflect the expected deterioration of the adiabatic
approximation with the excitation energy.

For the second example of $^{130}$Ba with two protons in the system,
the TW geometry exhibits greater stability and is sustained over
a much wider range of spin values. The flip regime is encountered
at the maximal observed angular momentum of $I=24$. For the less
stable case of $^{135}$Pr the experimental data cover the transition
from TW to LW.

The construction of the CH from the adiabatic energy can be directly
applied to the microscopic TAC calculations. One can
expect that the $n=0$, $n=1$ and, with some reservation, the $n=2$
wobbling excitations will be described by such an extension of the
TAC method. Work along these lines is planned.

\section*{Acknowledgements}

This work was supported by the National Natural Science Foundation of
China under Grant No.~12205103 and the US DoE Grant DE-FG02-95ER40934.



\begin{thebibliography}{68}
\expandafter\ifx\csname
natexlab\endcsname\relax\def\natexlab#1{#1}\fi
\expandafter\ifx\csname bibnamefont\endcsname\relax
  \def\bibnamefont#1{#1}\fi
\expandafter\ifx\csname bibfnamefont\endcsname\relax
  \def\bibfnamefont#1{#1}\fi
\expandafter\ifx\csname citenamefont\endcsname\relax
  \def\citenamefont#1{#1}\fi
\expandafter\ifx\csname url\endcsname\relax
  \def\url#1{\texttt{#1}}\fi
\expandafter\ifx\csname urlprefix\endcsname\relax\def\urlprefix{URL
}\fi \providecommand{\bibinfo}[2]{#2}
\providecommand{\eprint}[2][]{\url{#2}}

\bibitem[{\citenamefont{Bohr and Mottelson}(1975)}]{Bohr1975}
\bibinfo{author}{\bibfnamefont{A.}~\bibnamefont{Bohr}} \bibnamefont{and}
  \bibinfo{author}{\bibfnamefont{B.~R.} \bibnamefont{Mottelson}},
  \emph{\bibinfo{title}{Nuclear structure}}, vol.~\bibinfo{volume}{II}
  (\bibinfo{publisher}{Benjamin, New York}, \bibinfo{year}{1975}).

\bibitem[{\citenamefont{Frauendorf and D\"onau}(2014)}]{Frauendorf2014PRC}
\bibinfo{author}{\bibfnamefont{S.}~\bibnamefont{Frauendorf}} \bibnamefont{and}
  \bibinfo{author}{\bibfnamefont{F.}~\bibnamefont{D\"onau}},
  \bibinfo{journal}{Phys. Rev. C} \textbf{\bibinfo{volume}{89}},
  \bibinfo{pages}{014322} (\bibinfo{year}{2014}).

\bibitem[{\citenamefont{Chen and Frauendorf}(2022)}]{Q.B.Chen2022EPJA}
\bibinfo{author}{\bibfnamefont{Q.~B.} \bibnamefont{Chen}} \bibnamefont{and}
  \bibinfo{author}{\bibfnamefont{S.}~\bibnamefont{Frauendorf}},
  \bibinfo{journal}{Eur. Phys. J. A} \textbf{\bibinfo{volume}{58}},
  \bibinfo{pages}{75} (\bibinfo{year}{2022}).

\bibitem[{\citenamefont{Chen et~al.}(2017)\citenamefont{Chen, Chen, Luo, Meng,
  and Zhang}}]{F.Q.Chen2017PRC}
\bibinfo{author}{\bibfnamefont{F.~Q.} \bibnamefont{Chen}},
  \bibinfo{author}{\bibfnamefont{Q.~B.} \bibnamefont{Chen}},
  \bibinfo{author}{\bibfnamefont{Y.~A.} \bibnamefont{Luo}},
  \bibinfo{author}{\bibfnamefont{J.}~\bibnamefont{Meng}}, \bibnamefont{and}
  \bibinfo{author}{\bibfnamefont{S.~Q.} \bibnamefont{Zhang}},
  \bibinfo{journal}{Phys. Rev. C} \textbf{\bibinfo{volume}{96}},
  \bibinfo{pages}{051303(R)} (\bibinfo{year}{2017}).

\bibitem[{\citenamefont{Chen and Meng}(2018)}]{Q.B.Chen2018PRC_v1}
\bibinfo{author}{\bibfnamefont{Q.~B.} \bibnamefont{Chen}} \bibnamefont{and}
  \bibinfo{author}{\bibfnamefont{J.}~\bibnamefont{Meng}},
  \bibinfo{journal}{Phys. Rev. C} \textbf{\bibinfo{volume}{98}},
  \bibinfo{pages}{031303(R)} (\bibinfo{year}{2018}).

\bibitem[{\citenamefont{Streck et~al.}(2018)\citenamefont{Streck, Chen, Kaiser,
  and Mei\ss{}ner}}]{Streck2018PRC}
\bibinfo{author}{\bibfnamefont{E.}~\bibnamefont{Streck}},
  \bibinfo{author}{\bibfnamefont{Q.~B.} \bibnamefont{Chen}},
  \bibinfo{author}{\bibfnamefont{N.}~\bibnamefont{Kaiser}}, \bibnamefont{and}
  \bibinfo{author}{\bibfnamefont{U.-G.} \bibnamefont{Mei\ss{}ner}},
  \bibinfo{journal}{Phys. Rev. C} \textbf{\bibinfo{volume}{98}},
  \bibinfo{pages}{044314} (\bibinfo{year}{2018}).

\bibitem[{\citenamefont{Bringel et~al.}(2005)\citenamefont{Bringel, Hagemann,
  H\"{u}bel, Al-khatib, Bednarczyk, B\"{u}rger, Curien, Gangopadhyay, Herskind,
  Jensen et~al.}}]{Bringel2005EPJA}
\bibinfo{author}{\bibfnamefont{P.}~\bibnamefont{Bringel}},
  \bibinfo{author}{\bibfnamefont{G.~B.} \bibnamefont{Hagemann}},
  \bibinfo{author}{\bibfnamefont{H.}~\bibnamefont{H\"{u}bel}},
  \bibinfo{author}{\bibfnamefont{A.}~\bibnamefont{Al-khatib}},
  \bibinfo{author}{\bibfnamefont{P.}~\bibnamefont{Bednarczyk}},
  \bibinfo{author}{\bibfnamefont{A.}~\bibnamefont{B\"{u}rger}},
  \bibinfo{author}{\bibfnamefont{D.}~\bibnamefont{Curien}},
  \bibinfo{author}{\bibfnamefont{G.}~\bibnamefont{Gangopadhyay}},
  \bibinfo{author}{\bibfnamefont{B.}~\bibnamefont{Herskind}},
  \bibinfo{author}{\bibfnamefont{D.~R.} \bibnamefont{Jensen}},
  \bibnamefont{et~al.}, \bibinfo{journal}{Eur. Phys. J. A}
  \textbf{\bibinfo{volume}{24}}, \bibinfo{pages}{167} (\bibinfo{year}{2005}).

\bibitem[{\citenamefont{{\O}deg{\aa}rd
  et~al.}(2001)\citenamefont{{\O}deg{\aa}rd, Hagemann, Jensen, Bergstr{\"o}m,
  Herskind, Sletten, T{\"o}rm{\"a}nen, Wilson, Tj{\o}m, Hamamoto
  et~al.}}]{Odegaard2001PRL}
\bibinfo{author}{\bibfnamefont{S.~W.} \bibnamefont{{\O}deg{\aa}rd}},
  \bibinfo{author}{\bibfnamefont{G.~B.} \bibnamefont{Hagemann}},
  \bibinfo{author}{\bibfnamefont{D.~R.} \bibnamefont{Jensen}},
  \bibinfo{author}{\bibfnamefont{M.}~\bibnamefont{Bergstr{\"o}m}},
  \bibinfo{author}{\bibfnamefont{B.}~\bibnamefont{Herskind}},
  \bibinfo{author}{\bibfnamefont{G.}~\bibnamefont{Sletten}},
  \bibinfo{author}{\bibfnamefont{S.}~\bibnamefont{T{\"o}rm{\"a}nen}},
  \bibinfo{author}{\bibfnamefont{J.~N.} \bibnamefont{Wilson}},
  \bibinfo{author}{\bibfnamefont{P.~O.} \bibnamefont{Tj{\o}m}},
  \bibinfo{author}{\bibfnamefont{I.}~\bibnamefont{Hamamoto}},
  \bibnamefont{et~al.}, \bibinfo{journal}{Phys. Rev. Lett.}
  \textbf{\bibinfo{volume}{86}}, \bibinfo{pages}{5866} (\bibinfo{year}{2001}).

\bibitem[{\citenamefont{Jensen et~al.}(2002)\citenamefont{Jensen, Hagemann,
  Hamamoto, {\O}deg{\aa}rd, Herskind, Sletten, Wilson, Spohr, H{\"u}bel,
  Bringel et~al.}}]{Jensen2002PRL}
\bibinfo{author}{\bibfnamefont{D.~R.} \bibnamefont{Jensen}},
  \bibinfo{author}{\bibfnamefont{G.~B.} \bibnamefont{Hagemann}},
  \bibinfo{author}{\bibfnamefont{I.}~\bibnamefont{Hamamoto}},
  \bibinfo{author}{\bibfnamefont{S.~W.} \bibnamefont{{\O}deg{\aa}rd}},
  \bibinfo{author}{\bibfnamefont{B.}~\bibnamefont{Herskind}},
  \bibinfo{author}{\bibfnamefont{G.}~\bibnamefont{Sletten}},
  \bibinfo{author}{\bibfnamefont{J.~N.} \bibnamefont{Wilson}},
  \bibinfo{author}{\bibfnamefont{K.}~\bibnamefont{Spohr}},
  \bibinfo{author}{\bibfnamefont{H.}~\bibnamefont{H{\"u}bel}},
  \bibinfo{author}{\bibfnamefont{P.}~\bibnamefont{Bringel}},
  \bibnamefont{et~al.}, \bibinfo{journal}{Phys. Rev. Lett.}
  \textbf{\bibinfo{volume}{89}}, \bibinfo{pages}{142503}
  (\bibinfo{year}{2002}).

\bibitem[{\citenamefont{Sch\"{o}nwa{\ss}er
  et~al.}(2003)\citenamefont{Sch\"{o}nwa{\ss}er, H\"{u}bel, Hagemann,
  Bednarczyk, Benzoni, Bracco, Bringel, Chapman, Curien, Domscheit
  et~al.}}]{Schonwasser2003PLB}
\bibinfo{author}{\bibfnamefont{G.}~\bibnamefont{Sch\"{o}nwa{\ss}er}},
  \bibinfo{author}{\bibfnamefont{H.}~\bibnamefont{H\"{u}bel}},
  \bibinfo{author}{\bibfnamefont{G.~B.} \bibnamefont{Hagemann}},
  \bibinfo{author}{\bibfnamefont{P.}~\bibnamefont{Bednarczyk}},
  \bibinfo{author}{\bibfnamefont{G.}~\bibnamefont{Benzoni}},
  \bibinfo{author}{\bibfnamefont{A.}~\bibnamefont{Bracco}},
  \bibinfo{author}{\bibfnamefont{P.}~\bibnamefont{Bringel}},
  \bibinfo{author}{\bibfnamefont{R.}~\bibnamefont{Chapman}},
  \bibinfo{author}{\bibfnamefont{D.}~\bibnamefont{Curien}},
  \bibinfo{author}{\bibfnamefont{J.}~\bibnamefont{Domscheit}},
  \bibnamefont{et~al.}, \bibinfo{journal}{Phys. Lett. B}
  \textbf{\bibinfo{volume}{552}}, \bibinfo{pages}{9} (\bibinfo{year}{2003}).

\bibitem[{\citenamefont{Amro et~al.}(2003)\citenamefont{Amro, Ma, Hagemann,
  Diamond, Domscheit, Fallon, Gorgen, Herskind, H\"ubel, Jensen
  et~al.}}]{Amro2003PLB}
\bibinfo{author}{\bibfnamefont{H.}~\bibnamefont{Amro}},
  \bibinfo{author}{\bibfnamefont{W.~C.} \bibnamefont{Ma}},
  \bibinfo{author}{\bibfnamefont{G.~B.} \bibnamefont{Hagemann}},
  \bibinfo{author}{\bibfnamefont{R.~M.} \bibnamefont{Diamond}},
  \bibinfo{author}{\bibfnamefont{J.}~\bibnamefont{Domscheit}},
  \bibinfo{author}{\bibfnamefont{P.}~\bibnamefont{Fallon}},
  \bibinfo{author}{\bibfnamefont{A.}~\bibnamefont{Gorgen}},
  \bibinfo{author}{\bibfnamefont{B.}~\bibnamefont{Herskind}},
  \bibinfo{author}{\bibfnamefont{H.}~\bibnamefont{H\"ubel}},
  \bibinfo{author}{\bibfnamefont{D.~R.} \bibnamefont{Jensen}},
  \bibnamefont{et~al.}, \bibinfo{journal}{Phys. Lett. B}
  \textbf{\bibinfo{volume}{553}}, \bibinfo{pages}{197} (\bibinfo{year}{2003}).

\bibitem[{\citenamefont{Hartley et~al.}(2009)\citenamefont{Hartley, Janssens,
  Riedinger, Riley, Aguilar, Carpenter, Chiara, Chowdhury, Darby, Garg
  et~al.}}]{Hartley2009PRC}
\bibinfo{author}{\bibfnamefont{D.~J.} \bibnamefont{Hartley}},
  \bibinfo{author}{\bibfnamefont{R.~V.~F.} \bibnamefont{Janssens}},
  \bibinfo{author}{\bibfnamefont{L.~L.} \bibnamefont{Riedinger}},
  \bibinfo{author}{\bibfnamefont{M.~A.} \bibnamefont{Riley}},
  \bibinfo{author}{\bibfnamefont{A.}~\bibnamefont{Aguilar}},
  \bibinfo{author}{\bibfnamefont{M.~P.} \bibnamefont{Carpenter}},
  \bibinfo{author}{\bibfnamefont{C.~J.} \bibnamefont{Chiara}},
  \bibinfo{author}{\bibfnamefont{P.}~\bibnamefont{Chowdhury}},
  \bibinfo{author}{\bibfnamefont{I.~G.} \bibnamefont{Darby}},
  \bibinfo{author}{\bibfnamefont{U.}~\bibnamefont{Garg}}, \bibnamefont{et~al.},
  \bibinfo{journal}{Phys. Rev. C} \textbf{\bibinfo{volume}{80}},
  \bibinfo{pages}{041304(R)} (\bibinfo{year}{2009}).

\bibitem[{\citenamefont{Mukherjee et~al.}(2023)\citenamefont{Mukherjee,
  Bhattacharya, Trivedi, Tiwari, Singh, Muralithar, Yashraj, Katre, Kumar,
  Palit et~al.}}]{Mukherjee2023PRC}
\bibinfo{author}{\bibfnamefont{A.}~\bibnamefont{Mukherjee}},
  \bibinfo{author}{\bibfnamefont{S.}~\bibnamefont{Bhattacharya}},
  \bibinfo{author}{\bibfnamefont{T.}~\bibnamefont{Trivedi}},
  \bibinfo{author}{\bibfnamefont{S.}~\bibnamefont{Tiwari}},
  \bibinfo{author}{\bibfnamefont{R.~P.} \bibnamefont{Singh}},
  \bibinfo{author}{\bibfnamefont{S.}~\bibnamefont{Muralithar}},
  \bibinfo{author}{\bibnamefont{Yashraj}},
  \bibinfo{author}{\bibfnamefont{K.}~\bibnamefont{Katre}},
  \bibinfo{author}{\bibfnamefont{R.}~\bibnamefont{Kumar}},
  \bibinfo{author}{\bibfnamefont{R.}~\bibnamefont{Palit}},
  \bibnamefont{et~al.}, \bibinfo{journal}{Phys. Rev. C}
  \textbf{\bibinfo{volume}{107}}, \bibinfo{pages}{054310}
  (\bibinfo{year}{2023}).

\bibitem[{\citenamefont{Matta et~al.}(2015)\citenamefont{Matta, Garg, Li,
  Frauendorf, Ayangeakaa, Patel, Schlax, Palit, Saha, Sethi
  et~al.}}]{Matta2015PRL}
\bibinfo{author}{\bibfnamefont{J.~T.} \bibnamefont{Matta}},
  \bibinfo{author}{\bibfnamefont{U.}~\bibnamefont{Garg}},
  \bibinfo{author}{\bibfnamefont{W.}~\bibnamefont{Li}},
  \bibinfo{author}{\bibfnamefont{S.}~\bibnamefont{Frauendorf}},
  \bibinfo{author}{\bibfnamefont{A.~D.} \bibnamefont{Ayangeakaa}},
  \bibinfo{author}{\bibfnamefont{D.}~\bibnamefont{Patel}},
  \bibinfo{author}{\bibfnamefont{K.~W.} \bibnamefont{Schlax}},
  \bibinfo{author}{\bibfnamefont{R.}~\bibnamefont{Palit}},
  \bibinfo{author}{\bibfnamefont{S.}~\bibnamefont{Saha}},
  \bibinfo{author}{\bibfnamefont{J.}~\bibnamefont{Sethi}},
  \bibnamefont{et~al.}, \bibinfo{journal}{Phys. Rev. Lett.}
  \textbf{\bibinfo{volume}{114}}, \bibinfo{pages}{082501}
  (\bibinfo{year}{2015}).

\bibitem[{\citenamefont{Sensharma et~al.}(2019)\citenamefont{Sensharma, Garg,
  Zhu, Ayangeakaa, Frauendorf, Li, Bhat, Sheikh, Carpenter, Chen
  et~al.}}]{Sensharma2019PLB}
\bibinfo{author}{\bibfnamefont{N.}~\bibnamefont{Sensharma}},
  \bibinfo{author}{\bibfnamefont{U.}~\bibnamefont{Garg}},
  \bibinfo{author}{\bibfnamefont{S.}~\bibnamefont{Zhu}},
  \bibinfo{author}{\bibfnamefont{A.~D.} \bibnamefont{Ayangeakaa}},
  \bibinfo{author}{\bibfnamefont{S.}~\bibnamefont{Frauendorf}},
  \bibinfo{author}{\bibfnamefont{W.}~\bibnamefont{Li}},
  \bibinfo{author}{\bibfnamefont{G.}~\bibnamefont{Bhat}},
  \bibinfo{author}{\bibfnamefont{J.~A.} \bibnamefont{Sheikh}},
  \bibinfo{author}{\bibfnamefont{M.~P.} \bibnamefont{Carpenter}},
  \bibinfo{author}{\bibfnamefont{Q.~B.} \bibnamefont{Chen}},
  \bibnamefont{et~al.}, \bibinfo{journal}{Phys. Lett. B}
  \textbf{\bibinfo{volume}{792}}, \bibinfo{pages}{170 } (\bibinfo{year}{2019}).

\bibitem[{\citenamefont{Biswas et~al.}(2019)\citenamefont{Biswas, Palit,
  Frauendorf, Garg, Li, Bhat, Sheikh, Sethi, Saha, Singh
  et~al.}}]{Biswas2019EPJA}
\bibinfo{author}{\bibfnamefont{S.}~\bibnamefont{Biswas}},
  \bibinfo{author}{\bibfnamefont{R.}~\bibnamefont{Palit}},
  \bibinfo{author}{\bibfnamefont{S.}~\bibnamefont{Frauendorf}},
  \bibinfo{author}{\bibfnamefont{U.}~\bibnamefont{Garg}},
  \bibinfo{author}{\bibfnamefont{W.}~\bibnamefont{Li}},
  \bibinfo{author}{\bibfnamefont{G.~H.} \bibnamefont{Bhat}},
  \bibinfo{author}{\bibfnamefont{J.~A.} \bibnamefont{Sheikh}},
  \bibinfo{author}{\bibfnamefont{J.}~\bibnamefont{Sethi}},
  \bibinfo{author}{\bibfnamefont{S.}~\bibnamefont{Saha}},
  \bibinfo{author}{\bibfnamefont{P.}~\bibnamefont{Singh}},
  \bibnamefont{et~al.}, \bibinfo{journal}{Eur. Phys. J. A}
  \textbf{\bibinfo{volume}{55}}, \bibinfo{pages}{159} (\bibinfo{year}{2019}).

\bibitem[{\citenamefont{Sensharma et~al.}(2020)\citenamefont{Sensharma, Garg,
  Chen, Frauendorf, Burdette, Cozzi, Howard, Zhu, Carpenter, Copp
  et~al.}}]{Sensharma2020PRL}
\bibinfo{author}{\bibfnamefont{N.}~\bibnamefont{Sensharma}},
  \bibinfo{author}{\bibfnamefont{U.}~\bibnamefont{Garg}},
  \bibinfo{author}{\bibfnamefont{Q.~B.} \bibnamefont{Chen}},
  \bibinfo{author}{\bibfnamefont{S.}~\bibnamefont{Frauendorf}},
  \bibinfo{author}{\bibfnamefont{D.~P.} \bibnamefont{Burdette}},
  \bibinfo{author}{\bibfnamefont{J.~L.} \bibnamefont{Cozzi}},
  \bibinfo{author}{\bibfnamefont{K.~B.} \bibnamefont{Howard}},
  \bibinfo{author}{\bibfnamefont{S.}~\bibnamefont{Zhu}},
  \bibinfo{author}{\bibfnamefont{M.~P.} \bibnamefont{Carpenter}},
  \bibinfo{author}{\bibfnamefont{P.}~\bibnamefont{Copp}}, \bibnamefont{et~al.},
  \bibinfo{journal}{Phys. Rev. Lett.} \textbf{\bibinfo{volume}{124}},
  \bibinfo{pages}{052501} (\bibinfo{year}{2020}).

\bibitem[{\citenamefont{Nandi et~al.}(2020)\citenamefont{Nandi, Mukherjee,
  Chen, Frauendorf, Banik, Bhattacharya, Dar, Bhattacharyya, Bhattacharya,
  Chatterjee et~al.}}]{Nandi2020PRL}
\bibinfo{author}{\bibfnamefont{S.}~\bibnamefont{Nandi}},
  \bibinfo{author}{\bibfnamefont{G.}~\bibnamefont{Mukherjee}},
  \bibinfo{author}{\bibfnamefont{Q.~B.} \bibnamefont{Chen}},
  \bibinfo{author}{\bibfnamefont{S.}~\bibnamefont{Frauendorf}},
  \bibinfo{author}{\bibfnamefont{R.}~\bibnamefont{Banik}},
  \bibinfo{author}{\bibfnamefont{S.}~\bibnamefont{Bhattacharya}},
  \bibinfo{author}{\bibfnamefont{S.}~\bibnamefont{Dar}},
  \bibinfo{author}{\bibfnamefont{S.}~\bibnamefont{Bhattacharyya}},
  \bibinfo{author}{\bibfnamefont{C.}~\bibnamefont{Bhattacharya}},
  \bibinfo{author}{\bibfnamefont{S.}~\bibnamefont{Chatterjee}},
  \bibnamefont{et~al.}, \bibinfo{journal}{Phys. Rev. Lett.}
  \textbf{\bibinfo{volume}{125}}, \bibinfo{pages}{132501}
  (\bibinfo{year}{2020}).

\bibitem[{\citenamefont{Tim\'ar et~al.}(2019)\citenamefont{Tim\'ar, Chen,
  Kruzsicz, Sohler, Kuti, Zhang, Meng, Joshi, Wadsworth, Starosta
  et~al.}}]{Timar2019PRL}
\bibinfo{author}{\bibfnamefont{J.}~\bibnamefont{Tim\'ar}},
  \bibinfo{author}{\bibfnamefont{Q.~B.} \bibnamefont{Chen}},
  \bibinfo{author}{\bibfnamefont{B.}~\bibnamefont{Kruzsicz}},
  \bibinfo{author}{\bibfnamefont{D.}~\bibnamefont{Sohler}},
  \bibinfo{author}{\bibfnamefont{I.}~\bibnamefont{Kuti}},
  \bibinfo{author}{\bibfnamefont{S.~Q.} \bibnamefont{Zhang}},
  \bibinfo{author}{\bibfnamefont{J.}~\bibnamefont{Meng}},
  \bibinfo{author}{\bibfnamefont{P.}~\bibnamefont{Joshi}},
  \bibinfo{author}{\bibfnamefont{R.}~\bibnamefont{Wadsworth}},
  \bibinfo{author}{\bibfnamefont{K.}~\bibnamefont{Starosta}},
  \bibnamefont{et~al.}, \bibinfo{journal}{Phys. Rev. Lett.}
  \textbf{\bibinfo{volume}{122}}, \bibinfo{pages}{062501}
  (\bibinfo{year}{2019}).

\bibitem[{\citenamefont{Chakraborty et~al.}(2020)\citenamefont{Chakraborty,
  Sharma, Tiwary, Majumder, Gupta, Banerjee, Ganguly, Rai, Pragati, Mayank
  et~al.}}]{Chakraborty2020PLB}
\bibinfo{author}{\bibfnamefont{S.}~\bibnamefont{Chakraborty}},
  \bibinfo{author}{\bibfnamefont{H.~P.} \bibnamefont{Sharma}},
  \bibinfo{author}{\bibfnamefont{S.~S.} \bibnamefont{Tiwary}},
  \bibinfo{author}{\bibfnamefont{C.}~\bibnamefont{Majumder}},
  \bibinfo{author}{\bibfnamefont{A.~K.} \bibnamefont{Gupta}},
  \bibinfo{author}{\bibfnamefont{P.}~\bibnamefont{Banerjee}},
  \bibinfo{author}{\bibfnamefont{S.}~\bibnamefont{Ganguly}},
  \bibinfo{author}{\bibfnamefont{S.}~\bibnamefont{Rai}},
  \bibinfo{author}{\bibnamefont{Pragati}},
  \bibinfo{author}{\bibnamefont{Mayank}}, \bibnamefont{et~al.},
  \bibinfo{journal}{Phys. Lett. B} \textbf{\bibinfo{volume}{811}},
  \bibinfo{pages}{135854} (\bibinfo{year}{2020}).

\bibitem[{\citenamefont{{Rojeeta Devi} et~al.}(2021)\citenamefont{{Rojeeta
  Devi}, Kumar, Kumar, Neelam, Babra, Laskar, Biswas, Saha, Singh, Samanta
  et~al.}}]{Devi2021PLB}
\bibinfo{author}{\bibfnamefont{K.}~\bibnamefont{{Rojeeta Devi}}},
  \bibinfo{author}{\bibfnamefont{S.}~\bibnamefont{Kumar}},
  \bibinfo{author}{\bibfnamefont{N.}~\bibnamefont{Kumar}},
  \bibinfo{author}{\bibnamefont{Neelam}}, \bibinfo{author}{\bibfnamefont{F.~S.}
  \bibnamefont{Babra}}, \bibinfo{author}{\bibfnamefont{M.~S.~R.}
  \bibnamefont{Laskar}},
  \bibinfo{author}{\bibfnamefont{S.}~\bibnamefont{Biswas}},
  \bibinfo{author}{\bibfnamefont{S.}~\bibnamefont{Saha}},
  \bibinfo{author}{\bibfnamefont{P.}~\bibnamefont{Singh}},
  \bibinfo{author}{\bibfnamefont{S.}~\bibnamefont{Samanta}},
  \bibnamefont{et~al.}, \bibinfo{journal}{Phys. Lett. B}
  \textbf{\bibinfo{volume}{823}}, \bibinfo{pages}{136756}
  (\bibinfo{year}{2021}).

\bibitem[{\citenamefont{Petrache et~al.}(2019)\citenamefont{Petrache, Walker,
  Guo, Chen, Frauendorf, Liu, Wyss, Mengoni, Qiang, Astier
  et~al.}}]{Petrache2019PLB}
\bibinfo{author}{\bibfnamefont{C.~M.} \bibnamefont{Petrache}},
  \bibinfo{author}{\bibfnamefont{P.~M.} \bibnamefont{Walker}},
  \bibinfo{author}{\bibfnamefont{S.}~\bibnamefont{Guo}},
  \bibinfo{author}{\bibfnamefont{Q.~B.} \bibnamefont{Chen}},
  \bibinfo{author}{\bibfnamefont{S.}~\bibnamefont{Frauendorf}},
  \bibinfo{author}{\bibfnamefont{Y.~X.} \bibnamefont{Liu}},
  \bibinfo{author}{\bibfnamefont{R.~A.} \bibnamefont{Wyss}},
  \bibinfo{author}{\bibfnamefont{D.}~\bibnamefont{Mengoni}},
  \bibinfo{author}{\bibfnamefont{Y.~H.} \bibnamefont{Qiang}},
  \bibinfo{author}{\bibfnamefont{A.}~\bibnamefont{Astier}},
  \bibnamefont{et~al.}, \bibinfo{journal}{Phys. Lett. B}
  \textbf{\bibinfo{volume}{795}}, \bibinfo{pages}{241} (\bibinfo{year}{2019}).

\bibitem[{\citenamefont{Chen et~al.}(2019)\citenamefont{Chen, Frauendorf, and
  Petrache}}]{Q.B.Chen2019PRC_v1}
\bibinfo{author}{\bibfnamefont{Q.~B.} \bibnamefont{Chen}},
  \bibinfo{author}{\bibfnamefont{S.}~\bibnamefont{Frauendorf}},
  \bibnamefont{and} \bibinfo{author}{\bibfnamefont{C.~M.}
  \bibnamefont{Petrache}}, \bibinfo{journal}{Phys. Rev. C}
  \textbf{\bibinfo{volume}{100}}, \bibinfo{pages}{061301(R)}
  (\bibinfo{year}{2019}).

\bibitem[{\citenamefont{Wang et~al.}(2020)\citenamefont{Wang, Chen, and
  Zhao}}]{Y.K.Wang2020PLB}
\bibinfo{author}{\bibfnamefont{Y.~K.} \bibnamefont{Wang}},
  \bibinfo{author}{\bibfnamefont{F.~Q.} \bibnamefont{Chen}}, \bibnamefont{and}
  \bibinfo{author}{\bibfnamefont{P.~W.} \bibnamefont{Zhao}},
  \bibinfo{journal}{Phys. Lett. B} \textbf{\bibinfo{volume}{802}},
  \bibinfo{pages}{135246} (\bibinfo{year}{2020}).

\bibitem[{\citenamefont{Chen and Petrache}(2021)}]{F.Q.Chen2021PRC}
\bibinfo{author}{\bibfnamefont{F.-Q.} \bibnamefont{Chen}} \bibnamefont{and}
  \bibinfo{author}{\bibfnamefont{C.~M.} \bibnamefont{Petrache}},
  \bibinfo{journal}{Phys. Rev. C} \textbf{\bibinfo{volume}{103}},
  \bibinfo{pages}{064319} (\bibinfo{year}{2021}).

\bibitem[{\citenamefont{Lv et~al.}(2022{\natexlab{a}})\citenamefont{Lv,
  Petrache, Budaca, Astier, Zheng, Greenlees, Badran, Calverley, Cox, Grahn
  et~al.}}]{B.F.Lv2022PRC}
\bibinfo{author}{\bibfnamefont{B.~F.} \bibnamefont{Lv}},
  \bibinfo{author}{\bibfnamefont{C.~M.} \bibnamefont{Petrache}},
  \bibinfo{author}{\bibfnamefont{R.}~\bibnamefont{Budaca}},
  \bibinfo{author}{\bibfnamefont{A.}~\bibnamefont{Astier}},
  \bibinfo{author}{\bibfnamefont{K.~K.} \bibnamefont{Zheng}},
  \bibinfo{author}{\bibfnamefont{P.}~\bibnamefont{Greenlees}},
  \bibinfo{author}{\bibfnamefont{H.}~\bibnamefont{Badran}},
  \bibinfo{author}{\bibfnamefont{T.}~\bibnamefont{Calverley}},
  \bibinfo{author}{\bibfnamefont{D.~M.} \bibnamefont{Cox}},
  \bibinfo{author}{\bibfnamefont{T.}~\bibnamefont{Grahn}},
  \bibnamefont{et~al.}, \bibinfo{journal}{Phys. Rev. C}
  \textbf{\bibinfo{volume}{105}}, \bibinfo{pages}{034302}
  (\bibinfo{year}{2022}{\natexlab{a}}).

\bibitem[{\citenamefont{Frauendorf}(2018{\natexlab{a}})}]{Frauendorf2018PRC}
\bibinfo{author}{\bibfnamefont{S.}~\bibnamefont{Frauendorf}},
  \bibinfo{journal}{Phys. Rev. C} \textbf{\bibinfo{volume}{97}},
  \bibinfo{pages}{069801} (\bibinfo{year}{2018}{\natexlab{a}}).

\bibitem[{\citenamefont{Tanabe and Sugawara-Tanabe}(2018)}]{Tanabe2018PRC}
\bibinfo{author}{\bibfnamefont{K.}~\bibnamefont{Tanabe}} \bibnamefont{and}
  \bibinfo{author}{\bibfnamefont{K.}~\bibnamefont{Sugawara-Tanabe}},
  \bibinfo{journal}{Phys. Rev. C} \textbf{\bibinfo{volume}{97}},
  \bibinfo{pages}{069802} (\bibinfo{year}{2018}).

\bibitem[{\citenamefont{Lawrie et~al.}(2020)\citenamefont{Lawrie, Shirinda, and
  Petrache}}]{Lawrie2020PRC}
\bibinfo{author}{\bibfnamefont{E.~A.} \bibnamefont{Lawrie}},
  \bibinfo{author}{\bibfnamefont{O.}~\bibnamefont{Shirinda}}, \bibnamefont{and}
  \bibinfo{author}{\bibfnamefont{C.~M.} \bibnamefont{Petrache}},
  \bibinfo{journal}{Phys. Rev. C} \textbf{\bibinfo{volume}{101}},
  \bibinfo{pages}{034306} (\bibinfo{year}{2020}).

\bibitem[{\citenamefont{Lv et~al.}(2021)\citenamefont{Lv, Petrache, Lawrie,
  Astier, Dupont, Zheng, Greenlees, Badran, Calverley, Cox
  et~al.}}]{B.F.Lv2021PRC}
\bibinfo{author}{\bibfnamefont{B.~F.} \bibnamefont{Lv}},
  \bibinfo{author}{\bibfnamefont{C.~M.} \bibnamefont{Petrache}},
  \bibinfo{author}{\bibfnamefont{E.~A.} \bibnamefont{Lawrie}},
  \bibinfo{author}{\bibfnamefont{A.}~\bibnamefont{Astier}},
  \bibinfo{author}{\bibfnamefont{E.}~\bibnamefont{Dupont}},
  \bibinfo{author}{\bibfnamefont{K.~K.} \bibnamefont{Zheng}},
  \bibinfo{author}{\bibfnamefont{P.}~\bibnamefont{Greenlees}},
  \bibinfo{author}{\bibfnamefont{H.}~\bibnamefont{Badran}},
  \bibinfo{author}{\bibfnamefont{T.}~\bibnamefont{Calverley}},
  \bibinfo{author}{\bibfnamefont{D.~M.} \bibnamefont{Cox}},
  \bibnamefont{et~al.}, \bibinfo{journal}{Phys. Rev. C}
  \textbf{\bibinfo{volume}{103}}, \bibinfo{pages}{044308}
  (\bibinfo{year}{2021}).

\bibitem[{\citenamefont{Tanabe and Sugawara-Tanabe}(2017)}]{Tanabe2017PRC}
\bibinfo{author}{\bibfnamefont{K.}~\bibnamefont{Tanabe}} \bibnamefont{and}
  \bibinfo{author}{\bibfnamefont{K.}~\bibnamefont{Sugawara-Tanabe}},
  \bibinfo{journal}{Phys. Rev. C} \textbf{\bibinfo{volume}{95}},
  \bibinfo{pages}{064315} (\bibinfo{year}{2017}).

\bibitem[{\citenamefont{Lv et~al.}(2022{\natexlab{b}})\citenamefont{Lv,
  Petrache, Lawrie, Guo, Astier, Zheng, Ong, Wang, Zhou, Sun
  et~al.}}]{B.F.Lv2022PLB}
\bibinfo{author}{\bibfnamefont{B.~F.} \bibnamefont{Lv}},
  \bibinfo{author}{\bibfnamefont{C.~M.} \bibnamefont{Petrache}},
  \bibinfo{author}{\bibfnamefont{E.~A.} \bibnamefont{Lawrie}},
  \bibinfo{author}{\bibfnamefont{S.}~\bibnamefont{Guo}},
  \bibinfo{author}{\bibfnamefont{A.}~\bibnamefont{Astier}},
  \bibinfo{author}{\bibfnamefont{K.~K.} \bibnamefont{Zheng}},
  \bibinfo{author}{\bibfnamefont{H.~J.} \bibnamefont{Ong}},
  \bibinfo{author}{\bibfnamefont{J.~G.} \bibnamefont{Wang}},
  \bibinfo{author}{\bibfnamefont{X.~H.} \bibnamefont{Zhou}},
  \bibinfo{author}{\bibfnamefont{Z.~Y.} \bibnamefont{Sun}},
  \bibnamefont{et~al.}, \bibinfo{journal}{Phys. Lett. B}
  \textbf{\bibinfo{volume}{824}}, \bibinfo{pages}{136840}
  (\bibinfo{year}{2022}{\natexlab{b}}), ISSN \bibinfo{issn}{0370-2693}.

\bibitem[{\citenamefont{Guo et~al.}(2022)\citenamefont{Guo, Zhou, Petrache,
  Lawrie, Mthembu, Fang, Wu, Wang, Meng, Li et~al.}}]{S.Guo2022PLB}
\bibinfo{author}{\bibfnamefont{S.}~\bibnamefont{Guo}},
  \bibinfo{author}{\bibfnamefont{X.~H.} \bibnamefont{Zhou}},
  \bibinfo{author}{\bibfnamefont{C.~M.} \bibnamefont{Petrache}},
  \bibinfo{author}{\bibfnamefont{E.~A.} \bibnamefont{Lawrie}},
  \bibinfo{author}{\bibfnamefont{S.~H.} \bibnamefont{Mthembu}},
  \bibinfo{author}{\bibfnamefont{Y.~D.} \bibnamefont{Fang}},
  \bibinfo{author}{\bibfnamefont{H.~Y.} \bibnamefont{Wu}},
  \bibinfo{author}{\bibfnamefont{H.~L.} \bibnamefont{Wang}},
  \bibinfo{author}{\bibfnamefont{H.~Y.} \bibnamefont{Meng}},
  \bibinfo{author}{\bibfnamefont{G.~S.} \bibnamefont{Li}},
  \bibnamefont{et~al.}, \bibinfo{journal}{Phys. Lett. B}
  \textbf{\bibinfo{volume}{828}}, \bibinfo{pages}{137010}
  (\bibinfo{year}{2022}).

\bibitem[{\citenamefont{Nomura and Petrache}(2022)}]{Nomura2022PRC}
\bibinfo{author}{\bibfnamefont{K.}~\bibnamefont{Nomura}} \bibnamefont{and}
  \bibinfo{author}{\bibfnamefont{C.~M.} \bibnamefont{Petrache}},
  \bibinfo{journal}{Phys. Rev. C} \textbf{\bibinfo{volume}{105}},
  \bibinfo{pages}{024320} (\bibinfo{year}{2022}).

\bibitem[{\citenamefont{Shi and Chen}(2015)}]{W.X.Shi2015CPC}
\bibinfo{author}{\bibfnamefont{W.~X.} \bibnamefont{Shi}} \bibnamefont{and}
  \bibinfo{author}{\bibfnamefont{Q.~B.} \bibnamefont{Chen}},
  \bibinfo{journal}{Chin. Phys. C} \textbf{\bibinfo{volume}{39}},
  \bibinfo{pages}{054105} (\bibinfo{year}{2015}).

\bibitem[{\citenamefont{Qi et~al.}(2021)\citenamefont{Qi, Zhang, Wang, and
  Chen}}]{B.Qi2021JPG}
\bibinfo{author}{\bibfnamefont{B.}~\bibnamefont{Qi}},
  \bibinfo{author}{\bibfnamefont{H.}~\bibnamefont{Zhang}},
  \bibinfo{author}{\bibfnamefont{S.~Y.} \bibnamefont{Wang}}, \bibnamefont{and}
  \bibinfo{author}{\bibfnamefont{Q.~B.} \bibnamefont{Chen}},
  \bibinfo{journal}{J. Phys. G: Nucl. Part. Phys.}
  \textbf{\bibinfo{volume}{48}}, \bibinfo{pages}{055102}
  (\bibinfo{year}{2021}).

\bibitem[{\citenamefont{Hamamoto}(2002)}]{Hamamoto2002PRC}
\bibinfo{author}{\bibfnamefont{I.}~\bibnamefont{Hamamoto}},
  \bibinfo{journal}{Phys. Rev. C} \textbf{\bibinfo{volume}{65}},
  \bibinfo{pages}{044305} (\bibinfo{year}{2002}).

\bibitem[{\citenamefont{Hamamoto and Mottelson}(2003)}]{Hamamoto2003PRC}
\bibinfo{author}{\bibfnamefont{I.}~\bibnamefont{Hamamoto}} \bibnamefont{and}
  \bibinfo{author}{\bibfnamefont{B.~R.} \bibnamefont{Mottelson}},
  \bibinfo{journal}{Phys. Rev. C} \textbf{\bibinfo{volume}{68}},
  \bibinfo{pages}{034312} (\bibinfo{year}{2003}).

\bibitem[{\citenamefont{Chen et~al.}(2020)\citenamefont{Chen, Frauendorf,
  Kaiser, Mei{\ss}ner, and Meng}}]{Q.B.Chen2020PLB_v1}
\bibinfo{author}{\bibfnamefont{Q.~B.} \bibnamefont{Chen}},
  \bibinfo{author}{\bibfnamefont{S.}~\bibnamefont{Frauendorf}},
  \bibinfo{author}{\bibfnamefont{N.}~\bibnamefont{Kaiser}},
  \bibinfo{author}{\bibfnamefont{U.-G.} \bibnamefont{Mei{\ss}ner}},
  \bibnamefont{and} \bibinfo{author}{\bibfnamefont{J.}~\bibnamefont{Meng}},
  \bibinfo{journal}{Phys. Lett. B} \textbf{\bibinfo{volume}{807}},
  \bibinfo{pages}{135596} (\bibinfo{year}{2020}).

\bibitem[{\citenamefont{Broocks et~al.}(2021)\citenamefont{Broocks, Chen,
  Kaiser, and Mei\ss{}ner}}]{Broocks2021EPJA}
\bibinfo{author}{\bibfnamefont{C.}~\bibnamefont{Broocks}},
  \bibinfo{author}{\bibfnamefont{Q.~B.} \bibnamefont{Chen}},
  \bibinfo{author}{\bibfnamefont{N.}~\bibnamefont{Kaiser}}, \bibnamefont{and}
  \bibinfo{author}{\bibfnamefont{U.-G.} \bibnamefont{Mei\ss{}ner}},
  \bibinfo{journal}{Eur. Phys. J. A} \textbf{\bibinfo{volume}{57}},
  \bibinfo{pages}{161} (\bibinfo{year}{2021}).

\bibitem[{\citenamefont{Hu et~al.}(2021)\citenamefont{Hu, Peng, and
  Chen}}]{L.Hu2021PRC}
\bibinfo{author}{\bibfnamefont{L.}~\bibnamefont{Hu}},
  \bibinfo{author}{\bibfnamefont{J.}~\bibnamefont{Peng}}, \bibnamefont{and}
  \bibinfo{author}{\bibfnamefont{Q.~B.} \bibnamefont{Chen}},
  \bibinfo{journal}{Phys. Rev. C} \textbf{\bibinfo{volume}{104}},
  \bibinfo{pages}{064325} (\bibinfo{year}{2021}).

\bibitem[{\citenamefont{Zhang et~al.}(2022)\citenamefont{Zhang, Qi, Wang, Jia,
  and Wang}}]{H.Zhang2022PRC}
\bibinfo{author}{\bibfnamefont{H.}~\bibnamefont{Zhang}},
  \bibinfo{author}{\bibfnamefont{B.}~\bibnamefont{Qi}},
  \bibinfo{author}{\bibfnamefont{X.~D.} \bibnamefont{Wang}},
  \bibinfo{author}{\bibfnamefont{H.}~\bibnamefont{Jia}}, \bibnamefont{and}
  \bibinfo{author}{\bibfnamefont{S.~Y.} \bibnamefont{Wang}},
  \bibinfo{journal}{Phys. Rev. C} \textbf{\bibinfo{volume}{105}},
  \bibinfo{pages}{034339} (\bibinfo{year}{2022}).

\bibitem[{\citenamefont{Dai et~al.}(2023)\citenamefont{Dai, Chen, and
  Zhou}}]{H.M.Dai2023PRC}
\bibinfo{author}{\bibfnamefont{H.~M.} \bibnamefont{Dai}},
  \bibinfo{author}{\bibfnamefont{Q.~B.} \bibnamefont{Chen}}, \bibnamefont{and}
  \bibinfo{author}{\bibfnamefont{X.-R.} \bibnamefont{Zhou}},
  \bibinfo{journal}{Phys. Rev. C} \textbf{\bibinfo{volume}{108}},
  \bibinfo{pages}{054306} (\bibinfo{year}{2023}).

\bibitem[{\citenamefont{Raduta et~al.}(2017)\citenamefont{Raduta, Poenaru, and
  Ixaru}}]{Raduta2017PRC}
\bibinfo{author}{\bibfnamefont{A.~A.} \bibnamefont{Raduta}},
  \bibinfo{author}{\bibfnamefont{R.}~\bibnamefont{Poenaru}}, \bibnamefont{and}
  \bibinfo{author}{\bibfnamefont{L.~G.} \bibnamefont{Ixaru}},
  \bibinfo{journal}{Phys. Rev. C} \textbf{\bibinfo{volume}{96}},
  \bibinfo{pages}{054320} (\bibinfo{year}{2017}).

\bibitem[{\citenamefont{Budaca}(2018)}]{Budaca2018PRC}
\bibinfo{author}{\bibfnamefont{R.}~\bibnamefont{Budaca}},
  \bibinfo{journal}{Phys. Rev. C} \textbf{\bibinfo{volume}{97}},
  \bibinfo{pages}{024302} (\bibinfo{year}{2018}).

\bibitem[{\citenamefont{Budaca}(2021)}]{Budaca2021PRC}
\bibinfo{author}{\bibfnamefont{R.}~\bibnamefont{Budaca}},
  \bibinfo{journal}{Phys. Rev. C} \textbf{\bibinfo{volume}{103}},
  \bibinfo{pages}{044312} (\bibinfo{year}{2021}).

\bibitem[{\citenamefont{Budaca and Petrache}(2022)}]{Budaca2022PRC_v1}
\bibinfo{author}{\bibfnamefont{R.}~\bibnamefont{Budaca}} \bibnamefont{and}
  \bibinfo{author}{\bibfnamefont{C.~M.} \bibnamefont{Petrache}},
  \bibinfo{journal}{Phys. Rev. C} \textbf{\bibinfo{volume}{106}},
  \bibinfo{pages}{014313} (\bibinfo{year}{2022}).

\bibitem[{\citenamefont{Shimizu and Matsuzaki}(1995)}]{Shimizu1995NPA}
\bibinfo{author}{\bibfnamefont{Y.~R.} \bibnamefont{Shimizu}} \bibnamefont{and}
  \bibinfo{author}{\bibfnamefont{M.}~\bibnamefont{Matsuzaki}},
  \bibinfo{journal}{Nucl. Phys. A} \textbf{\bibinfo{volume}{588}},
  \bibinfo{pages}{559} (\bibinfo{year}{1995}).

\bibitem[{\citenamefont{Matsuzaki et~al.}(2002)\citenamefont{Matsuzaki,
  Shimizu, and Matsuyanagi}}]{Matsuzaki2002PRC}
\bibinfo{author}{\bibfnamefont{M.}~\bibnamefont{Matsuzaki}},
  \bibinfo{author}{\bibfnamefont{Y.~R.} \bibnamefont{Shimizu}},
  \bibnamefont{and}
  \bibinfo{author}{\bibfnamefont{K.}~\bibnamefont{Matsuyanagi}},
  \bibinfo{journal}{Phys. Rev. C} \textbf{\bibinfo{volume}{65}},
  \bibinfo{pages}{041303} (\bibinfo{year}{2002}).

\bibitem[{\citenamefont{Matsuzaki and Ohtsubo}(2004)}]{Matsuzaki2004PRCa}
\bibinfo{author}{\bibfnamefont{M.}~\bibnamefont{Matsuzaki}} \bibnamefont{and}
  \bibinfo{author}{\bibfnamefont{S.}~\bibnamefont{Ohtsubo}},
  \bibinfo{journal}{Phys. Rev. C} \textbf{\bibinfo{volume}{69}},
  \bibinfo{pages}{064317} (\bibinfo{year}{2004}).

\bibitem[{\citenamefont{Matsuzaki et~al.}(2004)\citenamefont{Matsuzaki,
  Shimizu, and Matsuyanagi}}]{Matsuzaki2004PRC}
\bibinfo{author}{\bibfnamefont{M.}~\bibnamefont{Matsuzaki}},
  \bibinfo{author}{\bibfnamefont{Y.~R.} \bibnamefont{Shimizu}},
  \bibnamefont{and}
  \bibinfo{author}{\bibfnamefont{K.}~\bibnamefont{Matsuyanagi}},
  \bibinfo{journal}{Phys. Rev. C} \textbf{\bibinfo{volume}{69}},
  \bibinfo{pages}{034325} (\bibinfo{year}{2004}).

\bibitem[{\citenamefont{Shimizu et~al.}(2008)\citenamefont{Shimizu, Shoji, and
  Matsuzaki}}]{Shimizu2008PRC}
\bibinfo{author}{\bibfnamefont{Y.~R.} \bibnamefont{Shimizu}},
  \bibinfo{author}{\bibfnamefont{T.}~\bibnamefont{Shoji}}, \bibnamefont{and}
  \bibinfo{author}{\bibfnamefont{M.}~\bibnamefont{Matsuzaki}},
  \bibinfo{journal}{Phys. Rev. C} \textbf{\bibinfo{volume}{77}},
  \bibinfo{pages}{024319} (\bibinfo{year}{2008}).

\bibitem[{\citenamefont{Shoji and Shimizu}(2009)}]{Shoji2009PTP}
\bibinfo{author}{\bibfnamefont{T.}~\bibnamefont{Shoji}} \bibnamefont{and}
  \bibinfo{author}{\bibfnamefont{Y.~R.} \bibnamefont{Shimizu}},
  \bibinfo{journal}{Progr. Theor. Phys.} \textbf{\bibinfo{volume}{121}},
  \bibinfo{pages}{319} (\bibinfo{year}{2009}).

\bibitem[{\citenamefont{Frauendorf and D\"onau}(2015)}]{Frauendorf2015PRC}
\bibinfo{author}{\bibfnamefont{S.}~\bibnamefont{Frauendorf}} \bibnamefont{and}
  \bibinfo{author}{\bibfnamefont{F.}~\bibnamefont{D\"onau}},
  \bibinfo{journal}{Phys. Rev. C} \textbf{\bibinfo{volume}{92}},
  \bibinfo{pages}{064306} (\bibinfo{year}{2015}).

\bibitem[{\citenamefont{Shimada et~al.}(2018)\citenamefont{Shimada, Fujioka,
  Tagami, and Shimizu}}]{Shimada2018PRC}
\bibinfo{author}{\bibfnamefont{M.}~\bibnamefont{Shimada}},
  \bibinfo{author}{\bibfnamefont{Y.}~\bibnamefont{Fujioka}},
  \bibinfo{author}{\bibfnamefont{S.}~\bibnamefont{Tagami}}, \bibnamefont{and}
  \bibinfo{author}{\bibfnamefont{Y.~R.} \bibnamefont{Shimizu}},
  \bibinfo{journal}{Phys. Rev. C} \textbf{\bibinfo{volume}{97}},
  \bibinfo{pages}{024318} (\bibinfo{year}{2018}).

\bibitem[{\citenamefont{Chen et~al.}(2014)\citenamefont{Chen, Zhang, Zhao, and
  Meng}}]{Q.B.Chen2014PRC}
\bibinfo{author}{\bibfnamefont{Q.~B.} \bibnamefont{Chen}},
  \bibinfo{author}{\bibfnamefont{S.~Q.} \bibnamefont{Zhang}},
  \bibinfo{author}{\bibfnamefont{P.~W.} \bibnamefont{Zhao}}, \bibnamefont{and}
  \bibinfo{author}{\bibfnamefont{J.}~\bibnamefont{Meng}},
  \bibinfo{journal}{Phys. Rev. C} \textbf{\bibinfo{volume}{90}},
  \bibinfo{pages}{044306} (\bibinfo{year}{2014}).

\bibitem[{\citenamefont{Chen et~al.}(2016{\natexlab{a}})\citenamefont{Chen,
  Zhang, and Meng}}]{Q.B.Chen2016PRC_v1}
\bibinfo{author}{\bibfnamefont{Q.~B.} \bibnamefont{Chen}},
  \bibinfo{author}{\bibfnamefont{S.~Q.} \bibnamefont{Zhang}}, \bibnamefont{and}
  \bibinfo{author}{\bibfnamefont{J.}~\bibnamefont{Meng}},
  \bibinfo{journal}{Phys. Rev. C} \textbf{\bibinfo{volume}{94}},
  \bibinfo{pages}{054308} (\bibinfo{year}{2016}{\natexlab{a}}).

\bibitem[{\citenamefont{Chen et~al.}(2013)\citenamefont{Chen, Zhang, Zhao,
  Jolos, and Meng}}]{Q.B.Chen2013PRC}
\bibinfo{author}{\bibfnamefont{Q.~B.} \bibnamefont{Chen}},
  \bibinfo{author}{\bibfnamefont{S.~Q.} \bibnamefont{Zhang}},
  \bibinfo{author}{\bibfnamefont{P.~W.} \bibnamefont{Zhao}},
  \bibinfo{author}{\bibfnamefont{R.~V.} \bibnamefont{Jolos}}, \bibnamefont{and}
  \bibinfo{author}{\bibfnamefont{J.}~\bibnamefont{Meng}},
  \bibinfo{journal}{Phys. Rev. C} \textbf{\bibinfo{volume}{87}},
  \bibinfo{pages}{024314} (\bibinfo{year}{2013}).

\bibitem[{\citenamefont{Chen et~al.}(2016{\natexlab{b}})\citenamefont{Chen,
  Zhang, Zhao, Jolos, and Meng}}]{Q.B.Chen2016PRC}
\bibinfo{author}{\bibfnamefont{Q.~B.} \bibnamefont{Chen}},
  \bibinfo{author}{\bibfnamefont{S.~Q.} \bibnamefont{Zhang}},
  \bibinfo{author}{\bibfnamefont{P.~W.} \bibnamefont{Zhao}},
  \bibinfo{author}{\bibfnamefont{R.~V.} \bibnamefont{Jolos}}, \bibnamefont{and}
  \bibinfo{author}{\bibfnamefont{J.}~\bibnamefont{Meng}},
  \bibinfo{journal}{Phys. Rev. C} \textbf{\bibinfo{volume}{94}},
  \bibinfo{pages}{044301} (\bibinfo{year}{2016}{\natexlab{b}}).

\bibitem[{\citenamefont{Wu et~al.}(2018)\citenamefont{Wu, Chen, Zhao, Zhang,
  and Meng}}]{X.H.Wu2018PRC}
\bibinfo{author}{\bibfnamefont{X.~H.} \bibnamefont{Wu}},
  \bibinfo{author}{\bibfnamefont{Q.~B.} \bibnamefont{Chen}},
  \bibinfo{author}{\bibfnamefont{P.~W.} \bibnamefont{Zhao}},
  \bibinfo{author}{\bibfnamefont{S.~Q.} \bibnamefont{Zhang}}, \bibnamefont{and}
  \bibinfo{author}{\bibfnamefont{J.}~\bibnamefont{Meng}},
  \bibinfo{journal}{Phys. Rev. C} \textbf{\bibinfo{volume}{98}},
  \bibinfo{pages}{064302} (\bibinfo{year}{2018}).

\bibitem[{\citenamefont{Kitagawa and Ueda}(1993)}]{Kitagawa1993PRA}
\bibinfo{author}{\bibfnamefont{M.}~\bibnamefont{Kitagawa}} \bibnamefont{and}
  \bibinfo{author}{\bibfnamefont{M.}~\bibnamefont{Ueda}},
  \bibinfo{journal}{Phys. Rev. A} \textbf{\bibinfo{volume}{47}},
  \bibinfo{pages}{5138} (\bibinfo{year}{1993}).

\bibitem[{\citenamefont{Klauder and Skagerstam}(1985)}]{Klauder1985book}
\bibinfo{author}{\bibfnamefont{J.}~\bibnamefont{Klauder}} \bibnamefont{and}
  \bibinfo{author}{\bibfnamefont{B.}~\bibnamefont{Skagerstam}},
  \emph{\bibinfo{title}{Coherent States: Applications in Physics and
  Mathematical Physics}} (\bibinfo{publisher}{World Scientific, Singapore},
  \bibinfo{year}{1985}).

\bibitem[{\citenamefont{Frauendorf}(2018{\natexlab{b}})}]{Frauendorf2018PS}
\bibinfo{author}{\bibfnamefont{S.}~\bibnamefont{Frauendorf}},
  \bibinfo{journal}{Phys. Scr.} \textbf{\bibinfo{volume}{93}},
  \bibinfo{pages}{043003} (\bibinfo{year}{2018}{\natexlab{b}}).

\bibitem[{\citenamefont{Podolsky}(1928)}]{Podolsky1928PR}
\bibinfo{author}{\bibfnamefont{B.}~\bibnamefont{Podolsky}},
  \bibinfo{journal}{Phys. Rev.} \textbf{\bibinfo{volume}{32}},
  \bibinfo{pages}{812} (\bibinfo{year}{1928}).

\bibitem[{\citenamefont{Pauli}(1933)}]{Pauli1933}
\bibinfo{author}{\bibfnamefont{W.}~\bibnamefont{Pauli}},
  \emph{\bibinfo{title}{in Handbuch der Physik}}, vol. \bibinfo{volume}{XXIV,
  p. 120} (\bibinfo{publisher}{Springer Verlag, Berlin}, \bibinfo{year}{1933}).

\bibitem[{\citenamefont{Frauendorf}(1993)}]{Frauendorf1993NPA}
\bibinfo{author}{\bibfnamefont{S.}~\bibnamefont{Frauendorf}},
  \bibinfo{journal}{Nucl. Phys. A} \textbf{\bibinfo{volume}{557}},
  \bibinfo{pages}{259c} (\bibinfo{year}{1993}).

\bibitem[{\citenamefont{Frauendorf}(2000)}]{Frauendorf2000NPA}
\bibinfo{author}{\bibfnamefont{S.}~\bibnamefont{Frauendorf}},
  \bibinfo{journal}{Nucl. Phys. A} \textbf{\bibinfo{volume}{677}},
  \bibinfo{pages}{115} (\bibinfo{year}{2000}).

\bibitem[{\citenamefont{Lewis et~al.}(1972)\citenamefont{Lewis, Malloy, JR.,
  Chao, and Laane}}]{Lewis1972JMS}
\bibinfo{author}{\bibfnamefont{J.~D.} \bibnamefont{Lewis}},
  \bibinfo{author}{\bibfnamefont{T.~B.} \bibnamefont{Malloy}},
  \bibinfo{author}{\bibnamefont{JR.}}, \bibinfo{author}{\bibfnamefont{T.~H.}
  \bibnamefont{Chao}}, \bibnamefont{and}
  \bibinfo{author}{\bibfnamefont{J.}~\bibnamefont{Laane}}, \bibinfo{journal}{J.
  Mod. Structure} \textbf{\bibinfo{volume}{12}}, \bibinfo{pages}{427}
  (\bibinfo{year}{1972}).

\end{thebibliography}

%

\end{document}